\documentclass[3p,times]{elsarticle}

\usepackage{textcomp}
\usepackage{amsmath,amssymb,amsfonts}
\usepackage{bbding}
\usepackage{arydshln}
\usepackage{colortbl}
\usepackage{multirow}
\usepackage[table,xcdraw]{xcolor}

\usepackage{upgreek}

\usepackage{amssymb}
\usepackage{amsmath}
\usepackage{multirow}
\usepackage[normalem]{ulem}
\useunder{\uline}{\ul}{}

\usepackage{algorithm}  
\usepackage{algorithmic}  

\usepackage[figuresright]{rotating}
\biboptions{sort&compress}

\begin{document}

\begin{frontmatter}




\title{An efficient dual-branch framework via implicit self-texture enhancement for arbitrary-scale histopathology image super-resolution}



\author[a,b]{Minghong Duan}

\author[a,b]{Linhao Qu}

\author[a,c]{Zhiwei Yang}

\author[a,b]{Manning Wang}

\author[a,b]{Chenxi Zhang\corref{cor1}}
\ead{chenxizhang@fudan.edu.cn}

\author[a,b]{Zhijian Song\corref{cor2}}
\ead{zjsong@fudan.edu.cn}

\cortext[cor1]{Corresponding author}
\cortext[cor2]{Principal corresponding author }

\address[a]{Digital Medical Research Center, School of Basic Medical Sciences, Fudan University, Shanghai 200032, China}
\address[b]{Shanghai Key Laboratory of Medical Image Computing and Computer Assisted Intervention, Shanghai 200032, China}
\address[c]{Academy for Engineering and Technology, Fudan University, Shanghai 200433, China}

\begin{abstract}
High-quality whole-slide scanning is expensive, complex, and time-consuming, thus limiting the acquisition and utilization of high-resolution histopathology images in daily clinical work. Deep learning-based single-image super-resolution (SISR) techniques provide an effective way to solve this problem. However, the existing SISR models applied in histopathology images can only work in fixed integer scaling factors, decreasing their applicability. Though methods based on implicit neural representation (INR) have shown promising results in arbitrary-scale super-resolution (SR) of natural images, applying them directly to histopathology images is inadequate because they have unique fine-grained image textures different from natural images. Thus, we propose an Implicit Self-Texture Enhancement-based dual-branch framework (ISTE) for arbitrary-scale SR of histopathology images to address this challenge. The proposed ISTE contains a feature aggregation branch and a texture learning branch. We employ the feature aggregation branch to enhance the learning of the local details for SR images while utilizing the texture learning branch to enhance the learning of high-frequency texture details. Then, we design a two-stage texture enhancement strategy to fuse the features from the two branches to obtain the SR images. Experiments on publicly available datasets, including TMA, HistoSR, and the TCGA lung cancer datasets, demonstrate that ISTE outperforms existing fixed-scale and arbitrary-scale SR algorithms across various scaling factors. Additionally, extensive experiments have shown that the histopathology images reconstructed by the proposed ISTE are applicable to downstream pathology image analysis tasks.
\end{abstract}

\begin{keyword}
super-resolution\sep histopathology images\sep implicit neural representation



\end{keyword}

\end{frontmatter}


\section{Introduction}
High-resolution (HR) whole slide images (WSIs) contain rich cellular morphology and pathological patterns, and they are the gold standard for clinical diagnosis and the basis for automated histopathology image analysis tasks, including segmentation and classification \cite{gilbertson2006primary,pantanowitz2011review,weinstein2004array,wilbur2011digital}. However, the acquisition and utilization of digital WSIs remain limited in the daily clinical workflow \cite{ghaznavi2013digital,wilbur2011digital}. On the one hand, HR digital WSIs are typically obtained through sophisticated and costly whole-slide scanning equipment, which is often difficult to access in remote and underserved regions. On the other hand, acquiring HR digital WSIs involves using dedicated micro-cameras within the whole slide scanner to capture image fragments from different local regions of the specimen, which are then stitched together to form a complete image depicting the entire specimen \cite{wu2023mmsrnet}. Such a digital process is highly time-consuming \cite{ghaznavi2013digital,wilbur2011digital}. Furthermore, HR digital WSIs are very large, often reaching gigapixels, which places additional demands on clinical funding support, professional training, ample data storage, and efficient data management \cite{nielsen2010virtual,pantanowitz2011review}. Therefore, if it is possible to scan low-resolution (LR) histopathology images with cheaper devices while designing algorithms that can produce WSIs maintaining high quality, the digitization process could be accelerated, and the clinical application of automated techniques to analyze histopathology images could be promoted \cite{ghaznavi2013digital,wilbur2011digital,madabhushi2016image}.

\begin{figure*}[h]
    \centering
    \includegraphics[width=0.95\textwidth]{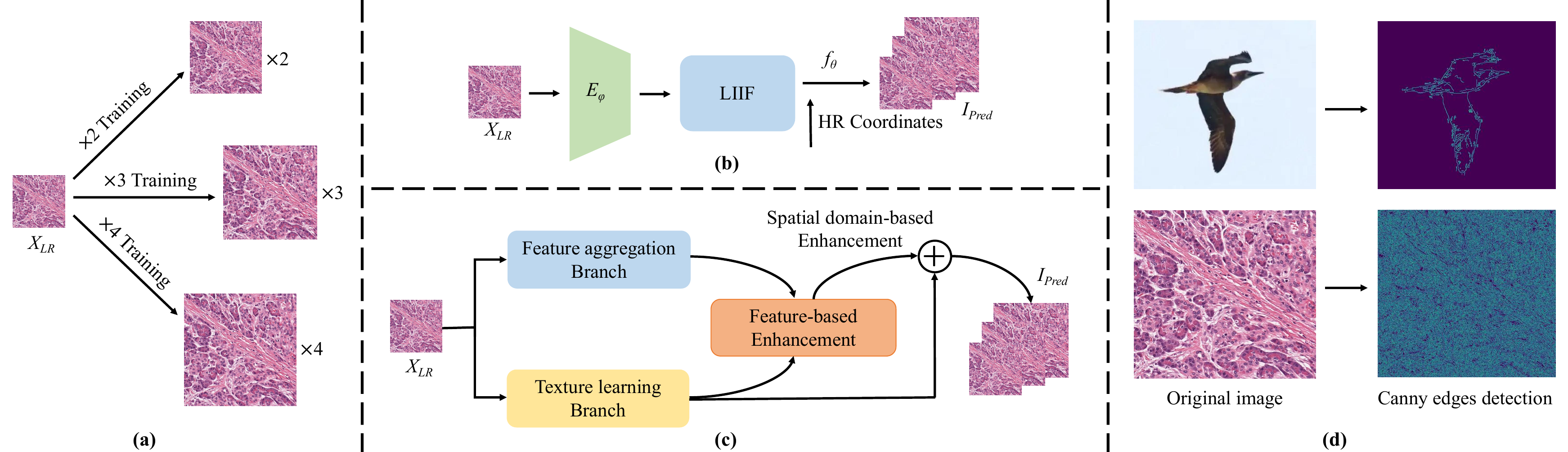}
    \caption{
        Motivation of our ISTE. 
        (a) Existing SR methods for histopathology images \cite{upadhyay2019mixed,mukherjee2018convolutional,juhong2023super,chen2020joint,shahidi2021breast,li2021single,wu2023mmsrnet} can only achieve fixed integer-scale SR and need to retrain the model to achieve different scaling factors; 
        (b) Existing SR algorithms based on implicit neural networks for natural images (exemplified by LIIF \cite{chen2021learning}) perform SR directly in the spatial domain, and lack attention and enhancement of image texture information; 
        (c) ISTE is an efficient dual-branch framework based on implicit self-texture enhancement for arbitrary-scale histopathology image SR. ISTE further enhances its performance through feature-based and spatial domain-based texture enhancement; 
        (d) We use the Canny operator \cite{canny1986computational} to extract texture from both natural and histopathology images. It is evident that, in contrast to natural images, histopathology images contain a large amount of fine-grained cell morphology and arrangement information, and they tend to have richer texture information.
    }
    \label{fig1}
\end{figure*}

Super-resolution (SR) algorithms based on deep learning can accurately map a single LR image to an HR image \cite{li2021single,lim2017enhanced,mukherjee2018convolutional}. Recently, deep learning-based methods have been widely applied in histopathology image SR. Most approaches construct a large dataset of LR-HR image pairs to train neural networks in an end-to-end manner. The trained neural networks can generate HR images with input LR images. For example, Mukherjee et al. \cite{mukherjee2018convolutional} utilized a convolutional neural network with an upsampling layer to produce SR images. Chen et al. \cite{chen2020joint} proposed a spatial wavelet dual-stream network to perform the SR image generation. As shown in Fig. \ref{fig1}(a), although these methods demonstrate commendable performance, they can only be trained and tested at a fixed integer scale, and the network needs to be retrained at a specific scale if other scaling factors are needed. However, in clinical pathological diagnosis, doctors usually need to continuously zoom in and out of sections at different scaling factors, so the applicability of these models is greatly limited. Unfortunately, to our knowledge, there are currently no models that can achieve arbitrary-scale SR for histopathology images.

Recently, inspired by implicit neural representation (INR) \cite{sitzmann2020implicit,tancik2020fourier,mildenhall2021nerf}, some studies have pioneered arbitrary-scale SR for natural images\cite{chen2021learning,lee2022local}. For example, Chen et al. \cite{chen2021learning} proposed the local implicit image function (LIIF), which represents 2D images as latent code through an encoder and maps the input coordinates and corresponding latent variables to RGB values through the decoding function based on the multilayer perceptron (MLP), enabling image SR at arbitrary scales. As shown in Fig. \ref{fig1}(b), although these methods can be directly applied to histopathology images, they do not account for the unique texture characteristics of histopathology images, resulting in sub-optimal performance. As shown in Fig. \ref{fig1}(d), histopathology images contain a large amount of fine-grained cell morphology and repetition, unlike natural images. Better reconstructing the unique texture characteristics at arbitrary scales is essential for histopathology image SR.

Motivated by the observation above, we propose an efficient dual-branch framework based on implicit self-texture enhancement (ISTE) for arbitrary-scale SR of histopathology images to better deal with its special texture. Fig. \ref{fig1}(c) briefly illustrates the overall framework of ISTE. Specifically, ISTE contains a feature aggregation branch and a texture learning branch. In the feature aggregation branch, we propose the local feature interactor (LFI) module, which is designed to enhance feature interaction within local regions and to sharpen the framework’s attention to image details. In the texture learning branch, we propose the texture learner (TL) to enhance the learning of high-frequency texture information. After that, we design a two-stage texture enhancement strategy for these two branches, where the first stage is feature-based texture enhancement, and the second stage is spatial domain-based texture enhancement. Considering that histopathology images contain many similar cell morphologies and periodic texture patterns, we assume that these similar regions can assist each other in reconstruction in the feature space, so we design the self-texture fusion (STF) module to accomplish feature-based texture enhancement. The main idea is to retrieve the texture information from the texture learning branch and transfer it to the feature aggregation branch for information fusion and enhancement. For spatial domain-based texture enhancement, we decode the features of the two branches into RGB values in the spatial domain using the local pixel decoder (LPD) and the local texture decoder (LTD), respectively, and perform information fusion in the spatial domain. These two decoders are based on implicit neural networks \cite{chen2021learning}, thus enabling image SR at arbitrary scales. Extensive experiments on three public datasets have shown that ISTE performs better than existing fixed-scale and arbitrary-scale SR algorithms at multiple scales and helps to improve downstream task performance. To the best of our knowledge, this is the first work to achieve arbitrary-scale SR in histopathology images. Overall, the contributions of this paper are as follows:

\begin{itemize}
\item {We introduce ISTE, an efficient dual-branch framework based on implicit self-texture enhancement for arbitrary-scale SR of histopathology images. ISTE recovers the texture details from the low resolution image through feature-based texture enhancement and spatial domain-based texture enhancement.}
\item {The proposed ISTE achieves state-of-the-art performance at various scaling factors on three public datasets, and we demonstrate the effectiveness of the proposed texture enhancement strategy through a series of ablation experiments.}
\item {The histopathology images reconstructed by ISTE are shown to be effective for two downstream tasks in pathology image analysis: gland segmentation and cancer detection. The performance of these tasks can be improved by using the reconstructed images.}
\end{itemize}

\section{Related Works}
\subsection{Deep learning-based super-resolution methods for natural images}
Single-image super-resolution (SISR) refers to recovering an HR image from an LR image or an LR image sequence, which is a classical low-level computer vision task with a wide range of applications. Deep neural networks can achieve accurate mapping from LR images to HR images due to their powerful fitting ability. Thus, they have become the mainstream approach in current SR studies. Numerous methods based on convolutional neural networks (CNNs) have been proposed for natural image SR, including SRCNN \cite{dong2014learning}, EDSR \cite{lim2017enhanced}, RDN \cite{zhang2018residual}, and RCAN \cite{zhang2018image}. To further improve the performance of SR, some methods utilized residual modules \cite{cavigelli2017cas,kim2016accurate}, densely connected modules \cite{wang2018esrgan,zhang2020residual}, and other blocks \cite{chen2016trainable,deng2021deep} for the design of the CNNs. Subsequently, a series of attention-based SR methods have emerged, such as channel attention \cite{niu2020single,zhang2018image}, self-attention (IPT \cite{chen2021pre}, SwinIR \cite{liang2021swinir}, HAT \cite{chen2023activating}), and non-local attention \cite{liu2018non,mei2021image}. However, these methods can only be trained and tested at a fixed integer scale, and the networks need to be retrained for new scaling factors.

In recent years, implicit neural representation (INR) has been proposed as a continuous data representation for various tasks in computer vision. INR uses a neural network (usually a coordinate-based MLP) to establish a mapping between coordinates and their signal values, which allows continuous and efficient modeling of 2D image signals. For example, Chen et al. \cite{chen2021learning} first used INR in the SR algorithm and proposed the local implicit image function (LIIF) for arbitrary-scale SR. Lee et al. \cite{lee2022local} proposed the local texture estimator (LTE), which transforms coordinates into Fourier domain information to enhance the representation of the local implicit function. Although these methods can be directly applied to pathological images for continuous magnification, they fail to recover the special textures in pathological images effectively.

\subsection{Deep learning-based super-resolution methods for pathological images}
In recent years, deep learning-based SR algorithms have been widely used in pathological images to improve imaging resolution \cite{upadhyay2019mixed,mukherjee2018convolutional,juhong2023super,chen2020joint,shahidi2021breast,li2021single,wu2023mmsrnet,ma2021stsrnet}. Upadhyay et al. \cite{upadhyay2019mixed} developed a generative adversarial network that considered pathological image SR and surgical smoke removal tasks at the same time. Mukherjee et al. \cite{mukherjee2018convolutional} implemented SR image generation using a CNN and up-sampling layer and augmented the outputs using the K-nearest neighbor algorithm. Chen et al. \cite{chen2020joint} accomplished the SR task through a spatial wavelet dual-stream network incorporating a refine context fusion module. Li et al. \cite{li2021single} utilized a generative adversarial network based on a multi-scale CNN for SR image generation and introduced a curriculum learning training strategy. Wu et al. \cite{wu2023mmsrnet} added a branch for magnification classification to the SR network and improved SR performance through multi-task learning. These studies demonstrate the promise of using SR to improve pathological image resolution in low-resource settings. However, they still have some limitations. For instance, they restrict training and testing to specific scaling factors, and the resultant SR outputs still exhibit scope for refinement. We attribute this primarily to a lack of adequate consideration for the unique textural characteristics of pathological images. In this paper, we introduce ISTE as a solution to overcome these challenges, aiming to achieve arbitrary-scale SR of pathological images with high quality.

\begin{figure}[h]
    \centering
    \includegraphics[width=0.95\textwidth]{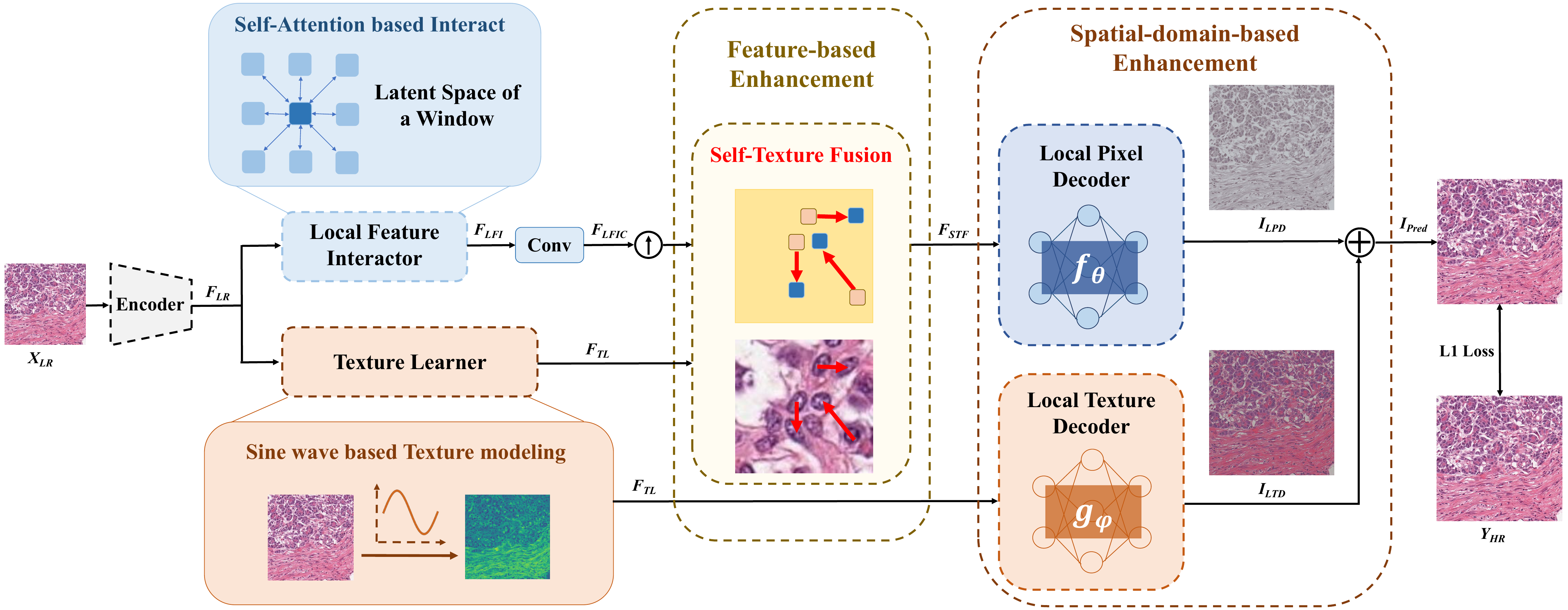}
    \caption{
Workflow of our ISTE. The LR image $X_{LR}$ is input into the encoder to get the pre-extracted feature map $F_{LR}$ first. In the feature aggregation branch, we input the feature $F_{LR}$ into the local feature interactor and a convolutional layer to obtain $F_{LFIC}$. In the texture learning branch, we input the feature $F_{LR}$ into the texture learner to obtain the texture feature $F_{TL}$. Then the feature maps from the two branches are input to the self-texture fusion module to accomplish feature-based enhancement. Finally, the enhanced feature $F_{STF}$ output from the STF module and the texture feature $F_{TL}$ output from the texture learner are decoded into RGB values respectively, and added up to accomplish spatial domain-based texture enhancement.
    }
    \label{fig2}
\end{figure}

\section{Method}
\subsection{Problem formulation and framework overview}
Given a set of $N$ pairs of corresponding LR images and HR images $\left\{X_{LR}^i, Y_{H R}^i\right\}_{i=1}^N$, the objective is to find the optimal parameters $\hat{\theta}$ of the SR model $F_\theta$:
\begin{equation}
\hat{\theta}=\arg _\theta \min \frac{1}{N} \sum_{i=1}^N L\left(F_\theta\left(X_{LR}^i\right), Y_{H R}^i\right)
\end{equation}
where $X_{LR}^i$ is a LR image and $Y_{HR}^i$ is its corresponding ground truth (GT), and L is the L1 loss function to measure the difference between the ground-truth and the generated HR images. 
Fig.2 shows the overall framework of our proposed ISTE. We first utilize SwinIR \cite{liang2021swinir} to perform feature pre-extraction on the input LR image $X_{LR}$ and then input the pre-extracted feature $F_{LR}$ into the upper feature aggregation branch and lower texture learning branch of ISTE, respectively. In the feature aggregation branch, we input the feature $F_{LR}$ into the local feature interactor (LFI) to enhance the interaction of features in the local region and obtain feature $F_{LFI}$. In the texture learning branch, we input the image feature $F_{LR}$ into the texture learner (TL) to enhance the learning of high-frequency information and extract the feature $F_{TL}$. Then we design a two-stage texture enhancement strategy for these two branches, where the first stage is feature-based texture enhancement, and the second stage is spatial domain-based texture enhancement. In the first stage, we designed the self-texture fusion (STF) module to leverage the interaction of similar regions of the pathological images in the feature space, thereby accomplishing feature-based texture enhancement to assist in reconstruction. In the second stage, we decode the $F_{STF}$ from the STF module to obtain the image $I_{LPD}$ through the local pixel decoder (LPD). Simultaneously, we decode the $F_{TL}$ from the TL module to obtain the image $I_{LTD}$ through the local texture decoder (LTD). Subsequently, we perform spatial summation of $I_{LTD}$ and $I_{LPD}$, obtaining the final reconstructed HR image $I_{Pred}$. The primary purpose of the second stage is to fully utilize the features $F_{TL}$ learned by the texture learner and decode them into the spatial domain for texture enhancement.

\subsection{Local feature interactor}
We propose the LFI to enhance the interaction of features within local regions, thereby capturing the correlation of features within local regions. As shown in Fig.3, the size of the feature map $F_{LR}$ is $h \times w \times 64$, and we denote each vector of $F_{LR}$ as $F_{LR}^j(j=1,2, \ldots, h \times w)$. The LFI first assigns a window of size $3 \times 3$ to each vector of $F_{LR}$, and the eight neighboring vectors in the window around $F_{LR}^j$ form a set $F_N^j=\left\{F_{N_i}^j \mid i=3,4, \ldots, 10\right\}$. The average pooling result of the vectors within a window is denoted as $F_{P}^j$. The feature map $F_{LFI}$ output by the LFI is calculated through self-attention so that each point on the feature map incorporates local features while paying more attention to itself. We denote each vector of $F_{LFI}$ as $F_{LFI}^j(j=1,2, \ldots, h \times w)$, and it is calculated through Eq.(2).
\begin{equation}
F_{L F I}^j=\sum_{i=1}^{10} \frac{\exp \left(\left(Q_{L R}^j\right)^T K_i^j\right)}{\sqrt{d} \Sigma_{i=1}^{10} \exp \left(\left(Q_{L R}^j\right)^T K_i^j\right)} V_i^j
\end{equation}
where $Q_{LR}^j$ is the query mapped linearly from $F_{LR}^j$, $K_1^j$ is the key mapped linearly from $F_{LR}^j$, $V_1^j$ is the value mapped linearly from $F_{LR}^j$, $K_2^j$ is the key mapped linearly from $F_{P}^j$, $V_2^j$ is the value mapped linearly from $F_{P}^j$, $\left\{K_i^j \mid i=3,4, \ldots, 10\right\}$ is the key mapped linearly from $F_{N}^j$, $\left\{V_i^j \mid i=3,4, \ldots, 10\right\}$ is the value mapped linearly from $F_{N}^j$, and $d$ is the dimension of these vectors. The parameters used by each window are shared in the self-attention calculation. 

\begin{figure}[h]
    \centering
    \includegraphics[width=0.55\textwidth]{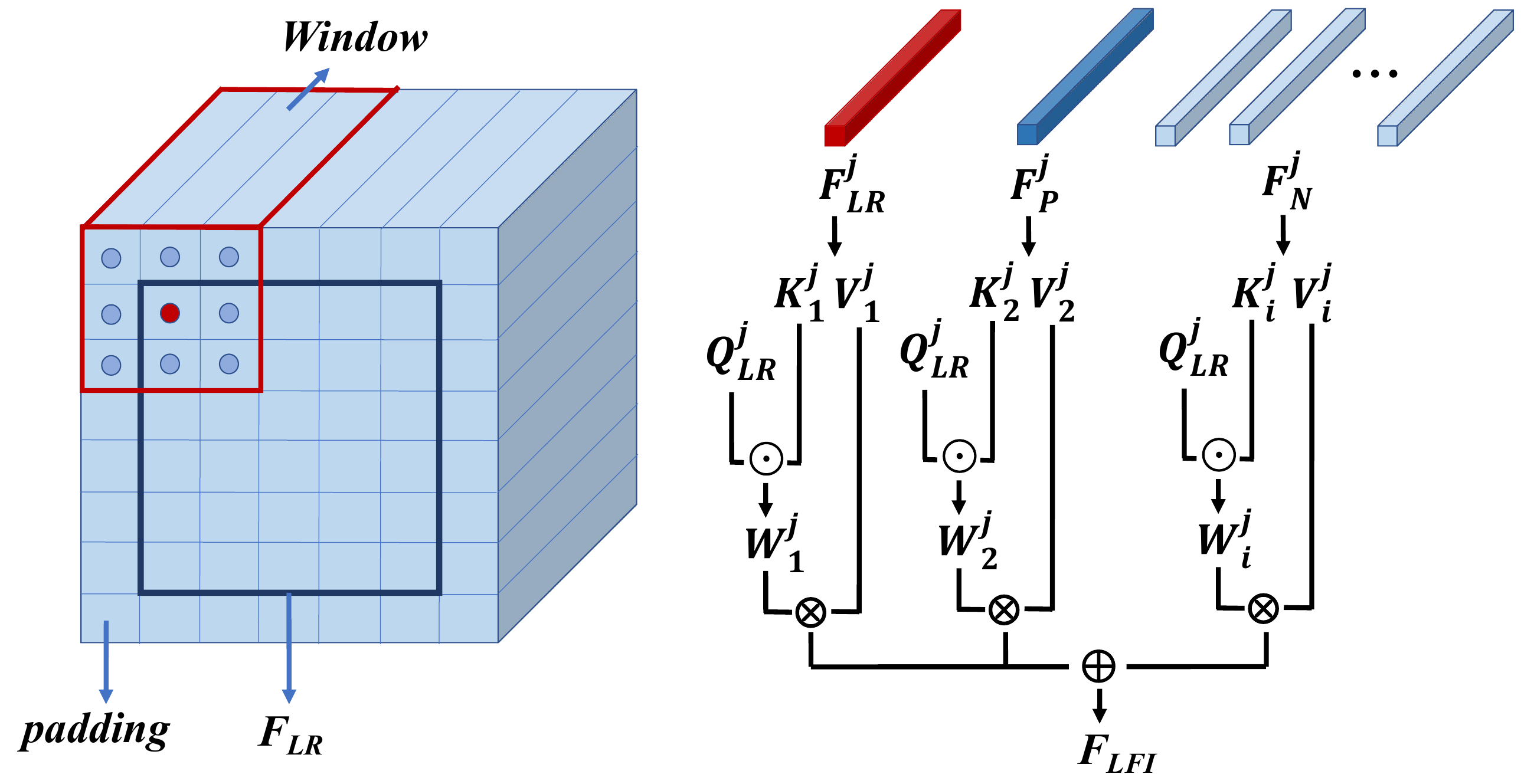}
    \caption{Local feature interactor.}
    \label{fig3}
\end{figure}

\subsection{Texture learner}
Inspired by LTE \cite{lee2022local}, we propose the TL for learning high-frequency texture information in pathological images. We employ sine activation to effectively enhance implicit neural representations for learning high-frequency details in images, thereby mitigating spectral bias issues stemming from the ReLU activation functions\cite{sitzmann2020implicit}. Specifically, we normalize the value of 2D pixel coordinate $\left(X^{\prime}, Y^{\prime}\right)=\left\{\left(\mathrm{x}_i^{\prime}, \mathrm{y}_j^{\prime}\right) \mid \mathrm{i}=1,2, \ldots, mw, \mathrm{j}=1,2, \ldots, mh\right\}$ in the continuous HR image domain and the value of 2D pixel coordinate $(X, Y)=\left\{\left(\mathrm{x}_i, \mathrm{y}_j\right) \mid \mathrm{i}=1,2, \ldots, mw, \mathrm{j}=1,2, \ldots, mh\right\}$ nearest to $\left(X^{\prime}, Y^{\prime}\right)$ in the continuous LR image domain between -1 and 1, and the Local Grid is defined as $\left(X^{\prime}-X, Y^{\prime}-Y\right)$. Since each pixel coordinate of the HR image has a corresponding coordinate in the LR image grid that is closest to it, the number of both the HR and LR image coordinates is equal to $mh\times mw$, where $m$ represents the scale factor. As shown in Fig.4(a), the TL module firstly outputs three feature maps $F_{Amp}\in h\times w\times256$, $F_{FreqX}\in h\times w\times256$ and $F_{FreqY}\in h\times w\times256$ through three $3\times3$ convolutional kernels respectively, and predicts the feature maps $Amp\in mh\times mw\times256$, $FreqX\in mh\times mw\times256$
and $FreqY\in mh\times mw\times256$ corresponding to each pixel coordinate of the HR image through nearest-neighbor interpolation. Then we use linear projection based on an MLP and Sigmoid activation function to map $(2 / \mathrm{mw}, 2 / \mathrm{mh})$ to a 256-dimensional feature vector $Phase$ to simulate the effect of texture fragment offset when the image scaling factor changes. The output of the TL module is calculated by Eq.(3):
\begin{equation}
\begin{aligned}
F_{TL}={Amp} \otimes & \operatorname{Sin}({FreqX} \odot \left(X^{\prime}-X\right)+ {FreqY} \odot \left(Y^{\prime}-Y\right)+{Phase})
\end{aligned}
\label{eq1}\end{equation}
where $\otimes$ represents element-wise multiplication and $\odot$ represents inner product operation.

\begin{figure}[h]
    \centering
    \includegraphics[width=0.75\textwidth]{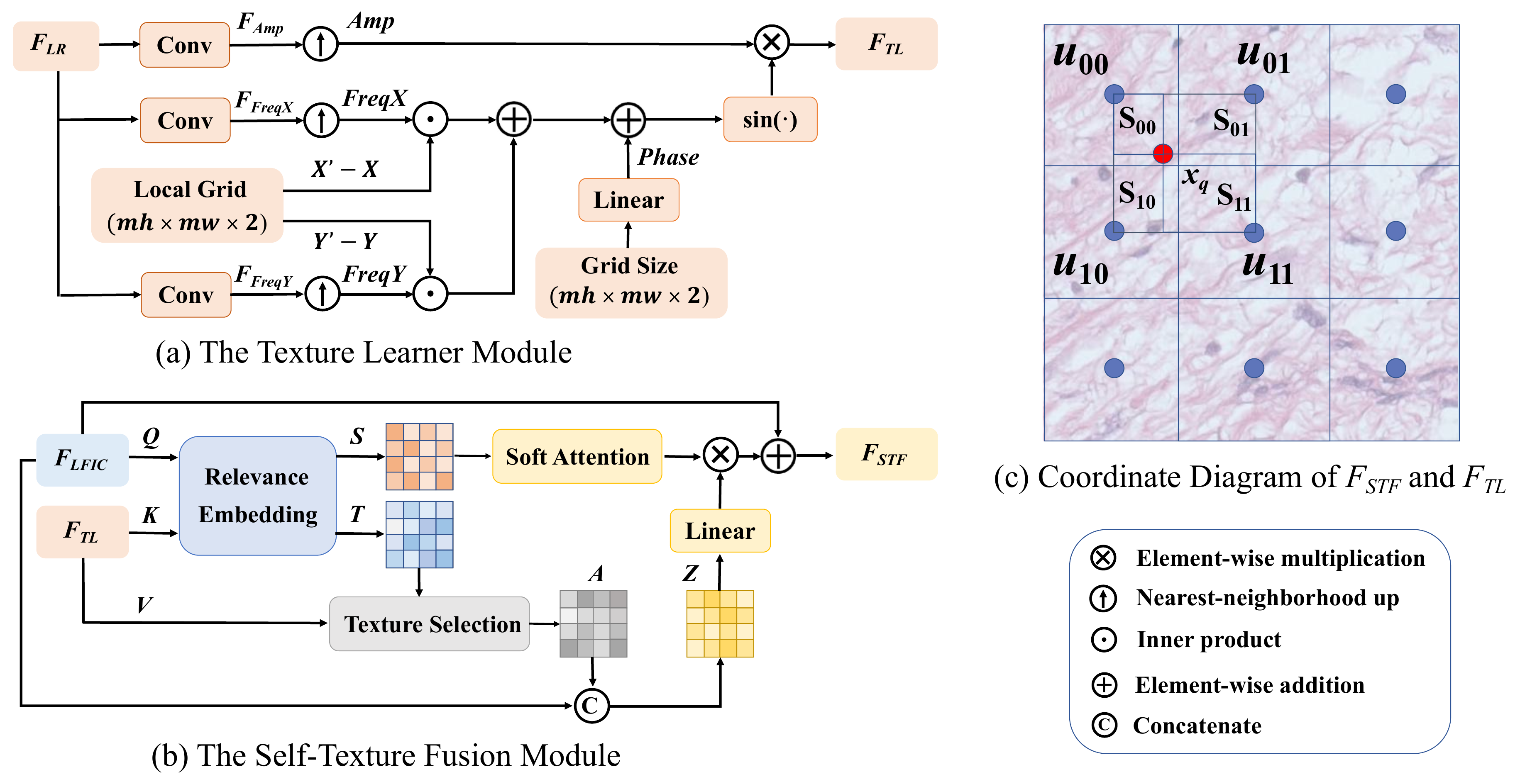}
    \caption{(a) Texture learner; (b) Self-texture fusion module; (c) Coordinate diagram of $F_{STF}$ and $F_{TL}$ for the local pixel decoder and local texture decoder.}
    \label{fig4}
\end{figure}

\subsection{Self-texture fusion module for feature-based enhancement}
Inspired by SRNTT \cite{zhang2019image} and T2Net \cite{feng2021task}, we propose the cross-attention-based STF module, whose main idea is to globally retrieve texture features most similar to $F_{LFIC}$ in $F_{TL}$ and fuse the retrieved features to $F_{LFIC}$, thus completing the feature-based texture enhancement. As shown in Fig.4(b), we use the features sampled from $F_{LFIC}$ by nearest-neighborhood interpolation as the query (Q) and use $F_{TL}$ as the key ($K$) and value ($V$) of the cross-attention module. To retrieve the texture features that are most relevant to the pixel feature $F_{LFIC}$, we first compute the similarity matrix $R$ of $Q$ and $K$, where each element $r_{i,j}$ of $R$ is computed according to Eq.(4), where $q_{i}$ represents an element of $Q$, and $k_{j}$ represents an element of $K$. Then we obtain the coordinate index matrix $T$ with the highest similarity to $q_{i}$ in $K$. An element in $T$ is $t_i=\arg \max _j\left(r_{i, j}\right)$, and $t_{i}$ represents the position coordinates of the texture feature $k_{j}$ with the highest similarity to $q_{i}$ in $F_{TL}$. We pick the feature vector $a_{i}$ with the highest similarity to each element in $Q$ from $V$ according to the coordinate index matrix $T$ to obtain the retrieved texture feature $A$, which can be represented by $a_i=v_{t_i}$ where $a_i$ is an element in $A$ and $v_{t_i}$ represents the element at the $t_i$-th position in $V$. To fuse the retrieved texture feature $A$ with the feature $F_{LFIC}$, we first concatenate $F_{LFIC}$ with $A$ and obtain the aggregated feature $Z$ through the output of an MLP, that is $Z=MLP(Concat(F_{LFIC}, A))$. Finally, we calculate the soft attention map $S$, where an element $s_i$ in $S$ represents the confidence of each element $a_i$ in the retrieved texture feature $A$, and $s_i=\max _j\left(r_{i, j}\right)$. $F_{STF}$ is calculated as below:
\begin{equation}
r_{i, j}=\left\langle\frac{q_i}{\left\|q_i\right\|}, \frac{k_j}{\left\|k_j\right\|}\right\rangle
\label{eq2}\end{equation}
\begin{equation}
F_{STF}=F_{LFIC} \oplus Z \otimes S
\label{eq3}\end{equation}
where $\langle\cdot\rangle$ represents inner product operation, $\|\cdot\|$ represents the square root operation, and $\oplus$ represents element-wise summation.

\subsection{Spatial domain-based enhancement}
In the spatial domain-based texture enhancement, we decode the texture feature $F_{TL}$ directly into the spatial domain $I_{LTD}$ and add it to $I_{LPD}$, which is reconstructed from $F_{FLIC}$ by the LPD, to obtain the final output $I_{Pred}$. Firstly, we utilize the LPD to decode the feature $F_{STF}$ into the RGB value $I_{LPD}$. We parameterize the LPD as an MLP. As shown in Fig.4(c), ${u_t}$ denotes the coordinates of the ${F_{LR}}$ and ${x_{q}}$ denotes the coordinates of the ${F_{STF}}$ and $F_{TL}$. We use $u_t(t \in 00,01,10,11)$ to denote the upper-left, upper-right, lower-left, and lower-right coordinates of an arbitrary point $x_{q}$, respectively. The RGB value at the coordinate ${x_{q}}$ in the HR image decoded by the LPD can be represented by Eq.(6), where $c$ contains two elements, $2/mh$ and $2/mw$, representing the sizes of each pixel in the $I_{LPD}$. Similarly, we calculate the RGB values of the texture information $I_{LTD}$ at coordinate $x_{q}$ via Eq.(7), where the LTD is parameterized as an MLP $g_{\varphi}$. We use the LTD to decode the texture features into the spatial domain texture information $I_{LTD}$ and add it to the $I_{LPD}$ via Eq.(8) for spatial domain texture enhancement to obtain the prediction result $I_{Pred}$, where ${\varphi}$ is the network parameter of the MLP $g_{\varphi}$. $S_t(t \in 00,01,10,11)$ is the area of the rectangular region between $x_q$ and $u_t$, and the weights are normalized by $S=\sum_{t \in\{00,01,10,11\}} S_t$.
\begin{equation}
I_{LPD}=\sum_{t \in\{00,01,10,11\}} \frac{S_t}{S} \cdot f_\theta\left(F_{STF}, x_q-u_t, {c}\right)
\label{eq4}\end{equation}
\begin{equation}
I_{LTD}=\sum_{t \in\{00,01,10,11\}} \frac{S_t}{S} \cdot g_{\varphi}\left(F_{T L}\right)
\label{eq5}\end{equation}
\begin{equation}
I_{Pred }=I_{LPD}+I_{LTD}
\label{eq6}\end{equation}

\section{Experiments}
We introduce the datasets, the implementation details, and the comparison to state-of-the-art SR methods in sections 4.1, 4.2, and 4.3, respectively. Then, we conduct a series of ablation studies in section 4.4. Finally, we perform two downstream task experiments, gland segmentation, and malignancy classification, to show that the HR images reconstructed by the proposed ISTE can help improve performance on downstream tasks in section 4.5.
\subsection{Datasets}
\subsubsection{Tissue Microarray (TMA) dataset}
Following Li et al. \cite{li2021single}, we experimented on the TMA dataset to validate our method. The TMA dataset, a widely used public dataset in pancreatic cancer research \cite{drifka2016highly,drifka2015periductal}, was scanned by an Aperio AT digital pathology scanner (Leica Biosystems, Wetzlar, Germany) at magnification of 0.504 $\upmu$m/pixel and contains 573 WSIs (average 3850×3850 pixels each). We randomly selected 460 WSIs as the training set, 57 WSIs as the validation set, and 56 WSIs as the test set.
\subsubsection{Histopathology Super-Resolution (HistoSR) dataset}
Following Chen et al. \cite{chen2020joint}, we conducted experiments on the Histopathology Super-Resolution (HistoSR) dataset, which is built on the high-quality H\&E stained WSIs of the Camelyon16 dataset. The HistoSR dataset contains HR images with a patch size of 192×192 through random cropping. The training set comprises 30000 HR patches, while the test set consists of 5000 HR patches. 
\subsubsection{TCGA Lung Cancer dataset}
The TCGA lung cancer dataset comprises 1054 WSIs (average 100000×100000 pixels each) \cite{li2021dual} from The Cancer Genome Atlas (TCGA) data center. We selected five slides from this dataset and cut them into 400 sub-images with a size of 3072×3072. We randomly selected 320 sub-images as the training set, 40 as the validation set, and 40 as the test set.
\subsection{Implementation details and evaluation metrics}
Following previous SR methods based on implicit neural representation \cite{chen2021learning,lee2022local}, we used the patches with the size of ${48 \times 48}$ as the input for training. We first randomly sampled the scaling factor ${m}$ in a uniform distribution U(1, 4) and cropped patches with the size of $48m \times 48m$ from the raw HR images in a batch, where ${m}$ represents the scaling factor. Following \cite{li2021single,ma2021stsrnet}, we resized the patches to ${48 \times 48}$ via bicubic downsampling and did a Gaussian blur to simulate degradation since it is difficult to acquire authentically downsampled images at arbitrary scales through scanners. The size of the Gaussian kernel was set to 1/2 of the scaling factor ${m}$. We sampled $48^2$ pixels from the corresponding cropped patches to form RGB-Coordinate pairs. We utilized the deep learning toolbox Pytorch to implement ISTE and Adam as the optimizer, setting the initial learning rate to 0.0001 and epochs to 1000. We employed structure similarity index measure (SSIM) and peak signal-to-noise ratio (PSNR) to evaluate the quality of reconstructed HR images.

\begin{figure}[h]
    \centering
    \includegraphics[width=0.95\textwidth]{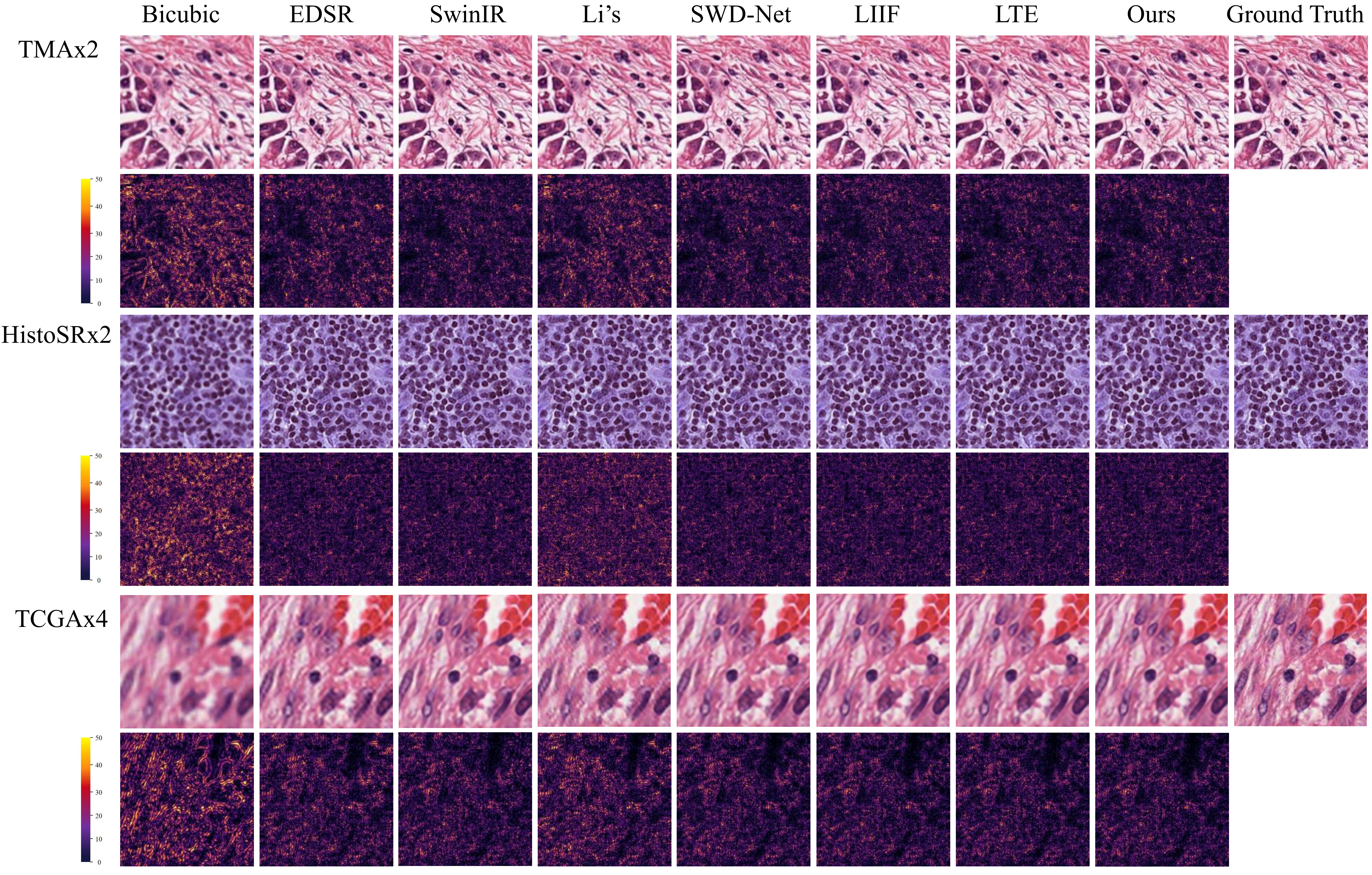}
    \caption{Visual comparison with error maps of different methods on the TMA, HistoSR, and TCGA datasets. The error map represents the absolute error value between the output result and the ground truth. The brighter the color, the greater the error.}
    \label{fig5}
\end{figure}

\subsection{Comparison with previous methods}
We compared the performance of ISTE with state-of-the-art SR methods in both the pathological image domain: SWD-Net \cite{chen2020joint} and Li et al. \cite{li2021single}, and the natural image domain: Bicubic, EDSR \cite{lim2017enhanced}, SwinIR \cite{liang2021swinir}, LIIF \cite{chen2021learning} and LTE \cite{lee2022local}, where the latter two are methods based on implicit neural representation. For a fair comparison, the backbone used for LIIF \cite{chen2021learning} and LTE \cite{lee2022local} is also SwinIR \cite{liang2021swinir} without upsampling layers.

\begin{table}[t]
    \centering
    \caption{Quantitative results of the proposed ISTE compared to state-of-the-art methods on the TMA, TCGA, and HistoSR datasets}
\setlength{\tabcolsep}{3pt}
\resizebox{1\columnwidth}{!}
    {
\begin{tabular}{|c|c|ccccccccc|cccccc|}
\hline
                          &                           & \multicolumn{9}{c|}{In-distribution}                                                                                                                                                                                                                                                                                                                                                                                                    & \multicolumn{6}{c|}{Out-of-distribution}                                                                                                                                                                                                                                           \\ \cline{3-17} 
                          &                           & \multicolumn{3}{c|}{×2}                                                                                                                            & \multicolumn{3}{c|}{×3}                                                                                                                            & \multicolumn{3}{c|}{×4}                                                                                                       & \multicolumn{3}{c|}{×6}                                                                                                                            & \multicolumn{3}{c|}{×8}                                                                                                       \\ \cline{3-17} 
\multirow{-3}{*}{Dataset} & \multirow{-3}{*}{Methods} & PSNR↑                                       & SSIM↑                                         & \multicolumn{1}{c|}{$P$ values}                        & PSNR↑                                       & SSIM↑                                         & \multicolumn{1}{c|}{$P$ values}                        & PSNR↑                                       & SSIM↑                                         & $P$ values                        & PSNR↑                                       & SSIM↑                                         & \multicolumn{1}{c|}{$P$ values}                        & PSNR↑                                       & SSIM↑                                         & $P$ values                        \\ \hline
                          & Bicubic                   & 28.54±2.890                                 & 0.8931±0.0474                                 & \multicolumn{1}{c|}{\textless 0.001/\textless 0.001} & 25.25±2.932                                 & 0.7708±0.1004                                 & \multicolumn{1}{c|}{\textless 0.001/\textless 0.001} & 23.43±2.915                                 & 0.6735±0.1407                                 & \textless 0.001/\textless 0.001 & 21.50±2.868                                 & 0.5647±0.1839                                 & \multicolumn{1}{c|}{\textless 0.001/\textless 0.001} & 20.44±2.849                                 & 0.5123±0.2042                                 & \textless 0.001/\textless 0.001 \\
                          & EDSR\cite{lim2017enhanced}                      & 30.54±2.792                                 & 0.9370±0.0272                                 & \multicolumn{1}{c|}{\textless 0.001/\textless 0.001} & 26.38±2.880                                 & 0.8228±0.0782                                 & \multicolumn{1}{c|}{\textless 0.001/\textless 0.001} & 24.94±2.884                                 & 0.7652±0.1014                                 & \textless 0.001/\textless 0.001 & -                                           & -                                             & \multicolumn{1}{c|}{-}                               & -                                           & -                                             & -                               \\
                          & SwinIR\cite{liang2021swinir}                    & 31.20±2.747                                 & {\color[HTML]{333333} 0.9438±0.0247}          & \multicolumn{1}{c|}{\textless 0.001/\textless 0.001} & 28.18±2.939                                 & 0.8773±0.0563                                 & \multicolumn{1}{c|}{\textless 0.001/\textless 0.001} & {\color[HTML]{343434} 26.26±2.954}          & {\color[HTML]{343434} 0.8092±0.0868}          & \textless 0.001/\textless 0.001 & -                                           & -                                             & \multicolumn{1}{c|}{-}                               & -                                           & -                                             & -                               \\
                          & Li et al.\cite{li2021single}                 & 29.50±2.754                                 & 0.9211±0.0334                                 & \multicolumn{1}{c|}{\textless 0.001/\textless 0.001} & 26.09±2.801                                 & 0.8207±0.0779                                 & \multicolumn{1}{c|}{\textless 0.001/\textless 0.001} & 24.06±2.770                                 & 0.7206±0.1211                                 & \textless 0.001/\textless 0.001 & -                                           & -                                             & \multicolumn{1}{c|}{-}                               & -                                           & -                                             & -                               \\
                          & SWD-Net\cite{chen2020joint}                   & 31.18±2.832                                 & 0.9430±0.0251                                 & \multicolumn{1}{c|}{\textless 0.001/\textless 0.001} & 28.06±2.946                                 & 0.8746±0.0574                                 & \multicolumn{1}{c|}{\textless 0.001/\textless 0.001} & 26.09±2.934                                 & 0.8024±0.0894                                 & \textless 0.001/\textless 0.001 & -                                           & -                                             & \multicolumn{1}{c|}{-}                               & -                                           & -                                             & -                               \\ \cline{2-17} 
                          & LIIF\cite{chen2021learning}                      & 30.76±2.562                                 & 0.9422±0.0253                                 & \multicolumn{1}{c|}{\textless 0.001/\textless 0.001} & 27.84±2.794                                 & 0.8745±0.0572                                 & \multicolumn{1}{c|}{\textless 0.001/\textless 0.001} & 25.87±2.858                                 & 0.7990±0.0908                                 & \textless 0.001/\textless 0.001 & 23.50±2.886                                 & 0.6751±0.1425                                 & \multicolumn{1}{c|}{\textless 0.001/\textless 0.001} & 22.05±2.874                                 & 0.5954±0.1741                                 & \textless 0.001/\textless 0.001 \\
                          & LTE\cite{lee2022local}                       & {\color[HTML]{343434} 31.26±2.834}          & {\color[HTML]{333333} 0.9434±0.0250}          & \multicolumn{1}{c|}{\textless 0.001/\textless 0.001} & {\color[HTML]{343434} 28.19±2.949}          & {\color[HTML]{343434} 0.8784±0.0558}          & \multicolumn{1}{c|}{\textless 0.001/\textless 0.001} & 26.22±2.975                                 & 0.8077±0.0875                                 & \textless 0.001/\textless 0.001 & {\color[HTML]{343434} 23.73±2.958}          & {\color[HTML]{343434} 0.6806±0.1409}          & \multicolumn{1}{c|}{\textless 0.001/\textless 0.001} & {\color[HTML]{343434} 22.17±2.926}          & {\color[HTML]{000000} \textbf{0.5974±0.1738}} & \textless 0.001/\textless 0.001 \\
\multirow{-8}{*}{TMA}     & ISTE(ours)                & {\color[HTML]{000000} \textbf{31.27±2.828}} & {\color[HTML]{000000} \textbf{0.9444±0.0243}} & \multicolumn{1}{c|}{-}                               & {\color[HTML]{000000} \textbf{28.23±2.954}} & {\color[HTML]{000000} \textbf{0.8809±0.0547}} & \multicolumn{1}{c|}{-}                               & {\color[HTML]{000000} \textbf{26.46±2.979}} & {\color[HTML]{000000} \textbf{0.8160±0.0842}} & -                               & {\color[HTML]{000000} \textbf{23.86±2.963}} & {\color[HTML]{000000} \textbf{0.6851±0.1393}} & \multicolumn{1}{c|}{-}                               & {\color[HTML]{000000} \textbf{22.19±2.931}} & {\color[HTML]{343434} 0.5965±0.1742}          & -                               \\ \hline
                          & Bicubic                   & 27.43±3.322                                 & 0.8585±0.0496                                 & \multicolumn{1}{c|}{\textless 0.001/\textless 0.001} & 23.88±3.394                                 & 0.6999±0.0936                                 & \multicolumn{1}{c|}{\textless 0.001/\textless 0.001} & 22.01±3.498                                 & 0.5770±0.1243                                 & \textless 0.001/\textless 0.001 & 19.95±3.654                                 & 0.4259±0.1678                                 & \multicolumn{1}{c|}{\textless 0.001/\textless 0.001} & 18.89±3.683                                 & 0.3529±0.1898                                 & \textless 0.001/\textless 0.001 \\
                          & EDSR\cite{lim2017enhanced}                      & 31.53±3.185                                 & {\color[HTML]{343434} 0.9407±0.0243}          & \multicolumn{1}{c|}{\textless 0.001/= 0.001}         & 27.81±3.261                                 & 0.8588±0.0559                                 & \multicolumn{1}{c|}{\textless 0.001/\textless 0.001} & 25.76±3.218                                 & 0.7820±0.0853                                 & \textless 0.001/\textless 0.001 & -                                           & -                                             & \multicolumn{1}{c|}{-}                               & -                                           & -                                             & -                               \\
                          & SwinIR\cite{liang2021swinir}                    & 31.51±3.213                                 & 0.9397±0.0243                                 & \multicolumn{1}{c|}{\textless 0.001/\textless 0.001} & 27.89±3.167                                 & 0.8624±0.0551                                 & \multicolumn{1}{c|}{\textless 0.001/\textless 0.001} & 25.90±3.213                                 & 0.7870±0.0822                                 & \textless 0.001/\textless 0.001 & -                                           & -                                             & \multicolumn{1}{c|}{-}                               & -                                           & -                                             & -                               \\
                          & Li et al.\cite{li2021single}                 & 28.98±3.133                                 & 0.9024±0.0360                                 & \multicolumn{1}{c|}{\textless 0.001/\textless 0.001} & 25.34±3.117                                 & 0.7843±0.0750                                 & \multicolumn{1}{c|}{\textless 0.001/\textless 0.001} & 23.50±3.164                                 & 0.6893±0.0992                                 & \textless 0.001/\textless 0.001 & -                                           & -                                             & \multicolumn{1}{c|}{-}                               & -                                           & -                                             & -                               \\
                          & SWD-Net\cite{chen2020joint}                   & 31.49±3.216                                 & 0.9393±0.0243                                 & \multicolumn{1}{c|}{\textless 0.001/\textless 0.001} & 27.87±3.253                                 & 0.8595±0.0559                                 & \multicolumn{1}{c|}{\textless 0.001/\textless 0.001} & 25.78±3.268                                 & 0.7810±0.0841                                 & \textless 0.001/\textless 0.001 & -                                           & -                                             & \multicolumn{1}{c|}{-}                               & -                                           & -                                             & -                               \\ \cline{2-17} 
                          & LIIF\cite{chen2021learning}                      & 31.56±3.212                                 & 0.9399±0.0243                                 & \multicolumn{1}{c|}{\textless 0.001/\textless 0.001} & {\color[HTML]{343434} 28.03±3.270}          & 0.8639±0.0549                                 & \multicolumn{1}{c|}{\textless 0.001/\textless 0.001} & {\color[HTML]{343434} 25.93±3.310}          & 0.7862±0.0820                                 & \textless 0.001/\textless 0.001 & 22.94±3.498                                 & 0.6279±0.1195                                 & \multicolumn{1}{c|}{\textless 0.001/\textless 0.001} & 20.87±3.821                                 & 0.4889±0.1598                                 & \textless 0.001/\textless 0.001 \\
                          & LTE\cite{lee2022local}                       & {\color[HTML]{343434} 31.58±3.244}          & 0.9403±0.0242                                 & \multicolumn{1}{c|}{\textless 0.001/\textless 0.001} & {\color[HTML]{343434} 28.03±3.286}          & {\color[HTML]{343434} 0.8647±0.0545}          & \multicolumn{1}{c|}{\textless 0.001/\textless 0.001} & {\color[HTML]{343434} 25.93±3.317}          & {\color[HTML]{343434} 0.7872±0.0816}          & \textless 0.001/\textless 0.001 & {\color[HTML]{343434} 22.95±3.500}          & {\color[HTML]{343434} 0.6298±0.1192}          & \multicolumn{1}{c|}{\textless 0.001/\textless 0.001} & {\color[HTML]{343434} 20.89±3.815}          & {\color[HTML]{343434} 0.4909±0.1588}          & \textless 0.001/\textless 0.001 \\
\multirow{-8}{*}{HistoSR} & ISTE(ours)                & {\color[HTML]{000000} \textbf{31.65±3.252}} & {\color[HTML]{000000} \textbf{0.9410±0.0239}} & \multicolumn{1}{c|}{-}                               & {\color[HTML]{000000} \textbf{28.14±3.299}} & {\color[HTML]{000000} \textbf{0.8673±0.0540}} & \multicolumn{1}{c|}{-}                               & {\color[HTML]{000000} \textbf{26.05±3.327}} & {\color[HTML]{000000} \textbf{0.7909±0.0813}} & -                               & {\color[HTML]{000000} \textbf{23.01±3.508}} & {\color[HTML]{000000} \textbf{0.6331±0.1186}} & \multicolumn{1}{c|}{-}                               & {\color[HTML]{000000} \textbf{20.94±3.828}} & {\color[HTML]{000000} \textbf{0.4948±0.1586}} & -                               \\ \hline
                          & Bicubic                   & 32.98±0.962                                 & 0.9353±0.0127                                 & \multicolumn{1}{c|}{\textless 0.001/\textless 0.001} & 28.12±0.858                                 & 0.8070±0.0271                                 & \multicolumn{1}{c|}{\textless 0.001/\textless 0.001} & 25.63±0.844                                 & 0.6874±0.0345                                 & \textless 0.001/\textless 0.001 & 23.05±0.873                                 & 0.5354±0.0401                                 & \multicolumn{1}{c|}{\textless 0.001/\textless 0.001} & 21.64±0.913                                 & 0.4606±0.0438                                 & \textless 0.001/\textless 0.001 \\
                          & EDSR\cite{lim2017enhanced}                      & 36.14±0.962                                 & 0.9709±0.0063                                 & \multicolumn{1}{c|}{\textless 0.001/\textless 0.001} & 31.16±0.914                                 & 0.9010±0.0183                                 & \multicolumn{1}{c|}{\textless 0.001/\textless 0.001} & 28.01±0.840                                 & 0.8074±0.0278                                 & \textless 0.001/\textless 0.001 & -                                           & -                                             & \multicolumn{1}{c|}{-}                               & -                                           & -                                             & -                               \\
                          & SwinIR\cite{liang2021swinir}                    & 36.73±0.971                                 & 0.9731±0.0058                                 & \multicolumn{1}{c|}{\textless 0.001/\textless 0.001} & 31.77±0.895                                 & 0.9094±0.0167                                 & \multicolumn{1}{c|}{\textless 0.001/\textless 0.001} & 28.83±0.813                                 & 0.8258±0.0251                                 & \textless 0.001/\textless 0.001 & -                                           & -                                             & \multicolumn{1}{c|}{-}                               & -                                           & -                                             & -                               \\
                          & Li et al.\cite{li2021single}                 & 34.61±0.842                                 & 0.9580±0.0073                                 & \multicolumn{1}{c|}{\textless 0.001/\textless 0.001} & 29.89±0.816                                 & 0.8725±0.0188                                 & \multicolumn{1}{c|}{\textless 0.001/\textless 0.001} & 26.57±0.769                                 & 0.7358±0.0280                                 & \textless 0.001/\textless 0.001 & -                                           & -                                             & \multicolumn{1}{c|}{-}                               & -                                           & -                                             & -                               \\
                          & SWD-Net\cite{chen2020joint}                   & 36.76±0.965                                 & 0.9734±0.0058                                 & \multicolumn{1}{c|}{\textless 0.001/\textless 0.001} & 31.73±0.914                                 & 0.9074±0.0172                                 & \multicolumn{1}{c|}{\textless 0.001/\textless 0.001} & 28.85±0.864                                 & 0.8219±0.0260                                 & \textless 0.001/\textless 0.001 & -                                           & -                                             & \multicolumn{1}{c|}{-}                               & -                                           & -                                             & -                               \\ \cline{2-17} 
                          & LIIF\cite{chen2021learning}                      & 36.92±0.957                                 & 0.9742±0.0055                                 & \multicolumn{1}{c|}{\textless 0.001/\textless 0.001} & {\color[HTML]{343434} 31.99±0.911}          & {\color[HTML]{343434} 0.9110±0.0163}          & \multicolumn{1}{c|}{\textless 0.001/\textless 0.001} & 29.08±0.866                                 & 0.8275±0.0251                                 & \textless 0.001/\textless 0.001 & {\color[HTML]{343434} 25.55±0.829}          & {\color[HTML]{343434} 0.6641±0.0349}          & \multicolumn{1}{c|}{\textless 0.001/\textless 0.001} & {\color[HTML]{343434} 23.72±0.859}          & {\color[HTML]{343434} 0.5609±0.0398}          & \textless 0.001/\textless 0.001 \\
                          & LTE\cite{lee2022local}                       & {\color[HTML]{343434} 36.99±0.975}          & {\color[HTML]{343434} 0.9748±0.0056}          & \multicolumn{1}{c|}{\textless 0.001/\textless 0.001} & 31.98±0.908                                 & 0.9109±0.0164                                 & \multicolumn{1}{c|}{\textless 0.001/\textless 0.001} & {\color[HTML]{343434} 29.11±0.866}          & {\color[HTML]{343434} 0.8280±0.0250}         & \textless 0.001/\textless 0.001 & 25.52±0.823                                 & 0.6617±0.0349                                 & \multicolumn{1}{c|}{\textless 0.001/\textless 0.001} & 23.67±0.853                                 & 0.5580±0.0398                                 & \textless 0.001/\textless 0.001 \\
\multirow{-8}{*}{TCGA}    & ISTE(ours)                & {\color[HTML]{000000} \textbf{37.76±1.034}} & {\color[HTML]{000000} \textbf{0.9796±0.0050}} & \multicolumn{1}{c|}{-}                               & {\color[HTML]{000000} \textbf{32.06±0.914}} & {\color[HTML]{000000} \textbf{0.9124±0.0163}} & \multicolumn{1}{c|}{-}                               & {\color[HTML]{000000} \textbf{29.19±0.867}} & {\color[HTML]{000000} \textbf{0.8307±0.0247}} & -                               & {\color[HTML]{000000} \textbf{25.61±0.821}} & {\color[HTML]{000000} \textbf{0.6674±0.0342}} & \multicolumn{1}{c|}{-}                               & {\color[HTML]{000000} \textbf{23.76±0.856}} & {\color[HTML]{000000} \textbf{0.5637±0.0395}} & -                               \\ \hline
\end{tabular}
    }
    \label{table1}
\end{table}

\begin{table}[h]
    \centering
    \caption{Quantitative results of the proposed ISTE compared to other arbitrary-scale SR methods on the TCGA datasets at non-integer scales.}
\setlength{\tabcolsep}{3pt}
\resizebox{1\columnwidth}{!}
    {
\begin{tabular}{|c|ccc|ccc|ccc|ccc|ccc|}
\hline
                       & \multicolumn{3}{c|}{×1.5}                                                                                                     & \multicolumn{3}{c|}{×2.4}                                                                                                     & \multicolumn{3}{c|}{×3.3}                                                                                                     & \multicolumn{3}{c|}{×4.2}                                                                                                     & \multicolumn{3}{c|}{×5.1}                                                                                                     \\ \cline{2-16} 
\multirow{-2}{*}{TCGA} & PSNR↑                                       & SSIM↑                                         & $P$ values                        & PSNR↑                                       & SSIM↑                                         & $P$ values                        & PSNR↑                                       & SSIM↑                                         & $P$ values                        & PSNR↑                                       & SSIM↑                                         & $P$ values                        & PSNR↑                                       & SSIM↑                                         & $P$ values                        \\ \hline
LIIF\cite{chen2021learning}                   & 42.95±0.938                                 & 0.9962±0.0010                                 & \textless 0.001/\textless 0.001 & 34.60±0.940                                 & {\color[HTML]{343434} 0.9532±0.0096}          & \textless 0.001/\textless 0.001 & {\color[HTML]{343434} 30.08±0.858}          & {\color[HTML]{343434} 0.8777±0.0197}          & \textless 0.001/\textless 0.001 & 27.92±0.832                                 & {\color[HTML]{343434} 0.8018±0.0266}          & \textless 0.001/\textless 0.001 & {\color[HTML]{343434} 26.64±0.821}          & {\color[HTML]{343434} 0.7285±0.0315}          & \textless 0.001/\textless 0.001 \\
LTE\cite{lee2022local}                    & {\color[HTML]{343434} 43.34±0.951}          & {\color[HTML]{343434} 0.9968±0.0009}          & \textless 0.001/\textless 0.001 & {\color[HTML]{343434} 34.61±0.943}          & {\color[HTML]{343434} 0.9532±0.0096}          & \textless 0.001/\textless 0.001 & {\color[HTML]{343434} 30.08±0.858}          & 0.8775±0.0197                                 & \textless 0.001/\textless 0.001 & {\color[HTML]{343434} 27.93±0.832}          & 0.8017±0.0266                                 & \textless 0.001/\textless 0.001 & 26.62±0.814                                 & 0.7267±0.0316                                 & \textless 0.001/\textless 0.001 \\
ISTE(ours)                   & {\color[HTML]{000000} \textbf{44.46±0.895}} & {\color[HTML]{000000} \textbf{0.9982±0.0006}} & -                               & {\color[HTML]{000000} \textbf{34.91±0.985}} & {\color[HTML]{000000} \textbf{0.9568±0.0094}} & -                               & {\color[HTML]{000000} \textbf{30.14±0.859}} & {\color[HTML]{000000} \textbf{0.8791±0.0196}} & -                               & {\color[HTML]{000000} \textbf{28.02±0.834}} & {\color[HTML]{000000} \textbf{0.8053±0.0263}} & -                               & {\color[HTML]{000000} \textbf{26.71±0.815}} & {\color[HTML]{000000} \textbf{0.7312±0.0309}} & -                               \\ \hline
\end{tabular}
    }
    \label{table2}
\end{table}

\subsubsection{Quantitative results}
We compared our ISTE with competitors at five scaling factors of $\times2$, $\times3$, $\times4$, $\times6$, and $\times8$. As shown in Table 1, our ISTE achieved the highest performance in terms of PSNR and SSIM metrics at each scaling factor on the HistoSR and TCGA datasets. Although our method’s SSIM metric at $\times8$ is slightly lower than LTE by 0.0009 on the TMA dataset, it outperforms the comparison method in PSNR metrics at all scaling factors and SSIM metrics at other scaling factors. We evaluate the significant difference between our ISTE and other methods using paired student's t-tests. ${P\textless0.001}$ was considered as a statistically significant level. We report the specific value for p-values a little bit larger than 0.001, while those smaller than 0.001 are not given a specific value. As can be seen from the p-values in Table 1, there is a statistically significant difference with p-values smaller than 0.001 in almost all cases.
To further assess the advantages of our method over other arbitrary scale SR Methods, we present comparative results in Table 2 for ISTE, LTE [14], and LIIF [32] at non-integer scaling factors. Our method demonstrates superior performance in terms of both PSNR and SSIM metrics. We also provide the Frechet Inception Distance (FID) score metric to evaluate the perceptual quality of images generated by different methods in Table 3. The results indicate that the textures of images generated by our method are more realistic, yielding perceptual effects superior to other arbitrary-scale SR methods. Please refer to supplementary materials for more comparisons.

\begin{table}[t!]
    \centering
    \caption{FID scores between the reconstructed images and the ground truth HR images.}
\setlength{\tabcolsep}{3pt}
\resizebox{0.3\columnwidth}{!}
    {
\begin{tabular}{|c|c|ccccc|}
\hline
                          &                           & \multicolumn{5}{c|}{FID score$\downarrow$}                                                                                                                                                                         \\ \cline{3-7} 
\multirow{-2}{*}{Dataset} & \multirow{-2}{*}{Methods} & ×2                                   & ×3                                    & ×4                                    & ×6                                     & ×8                                     \\ \hline
                          & LIIF\cite{chen2021learning}                      & 1.23                                 & 2.92                                  & 17.58                                 & 88.64                                  & {\color[HTML]{000000} \textbf{120.62}} \\
                          & LTE\cite{lee2022local}                       & 1.22                                 & 2.96                                  & 17.22                                 & 90.37                                  & 124.30                                  \\
\multirow{-3}{*}{TCGA}    & ISTE(ours)                      & {\color[HTML]{000000} \textbf{1.07}} & {\color[HTML]{000000} \textbf{2.86}}  & {\color[HTML]{000000} \textbf{16.45}} & {\color[HTML]{000000} \textbf{88.62}}  & 122.12                                 \\ \hline
                          & LIIF\cite{chen2021learning}                      & 3.63                                 & 6.11                                  & 17.14                                 & 53.55                                  & 82.50                                   \\
                          & LTE\cite{lee2022local}                       & 3.15                                 & 5.39                                  & 15.40                                  & 53.66                                  & 82.74                                  \\
\multirow{-3}{*}{TMA}     & ISTE(ours)                      & {\color[HTML]{000000} \textbf{2.77}} & {\color[HTML]{000000} \textbf{4.74}}  & {\color[HTML]{000000} \textbf{13.53}} & {\color[HTML]{000000} \textbf{49.27}}  & {\color[HTML]{000000} \textbf{75.32}}  \\ \hline
                          & LIIF\cite{chen2021learning}                      & 9.24                                 & 39.00                                    & 76.69                                 & 130.53                                 & 156.85                                 \\
                          & LTE\cite{lee2022local}                       & 9.54                                 & 39.05                                 & 77.06                                 & 130.56                                 & 154.28                                 \\
\multirow{-3}{*}{HistoSR} & ISTE(ours)                      & {\color[HTML]{000000} \textbf{8.92}} & {\color[HTML]{000000} \textbf{37.82}} & {\color[HTML]{000000} \textbf{75.45}} & {\color[HTML]{000000} \textbf{128.81}} & {\color[HTML]{000000} \textbf{153.27}} \\ \hline
\end{tabular}
    }
    \label{table3}
\end{table}

\subsubsection{Qualitative results}
Fig.5 shows the visual results and absolute error maps of different methods on the TCGA datasets at the scale of ×4, TMA datasets at the scale of ×2, and HistoSR datasets at the scale of ×2. Our proposed method performs better in restoring texture information, closely approximating the ground truth. Based on the brightness levels in the absolute error maps, it is observable that our method's error maps contain more dark regions, indicating more minor errors in the reconstructed results compared to other methods. Fig.6 shows an SR example of a comparison of LIIF and our ISTE at non-integer scales. It can be seen that ISTE achieves arbitrary-scale SR with clear cell structure and texture. As shown in the red box, two cells are connected due to blurring in the image generated by LIIF while they are still separated in the image generated by ISTE at the scale of ×7.3. 

\begin{figure}[t]
\centering
\includegraphics[width=0.6\textwidth]{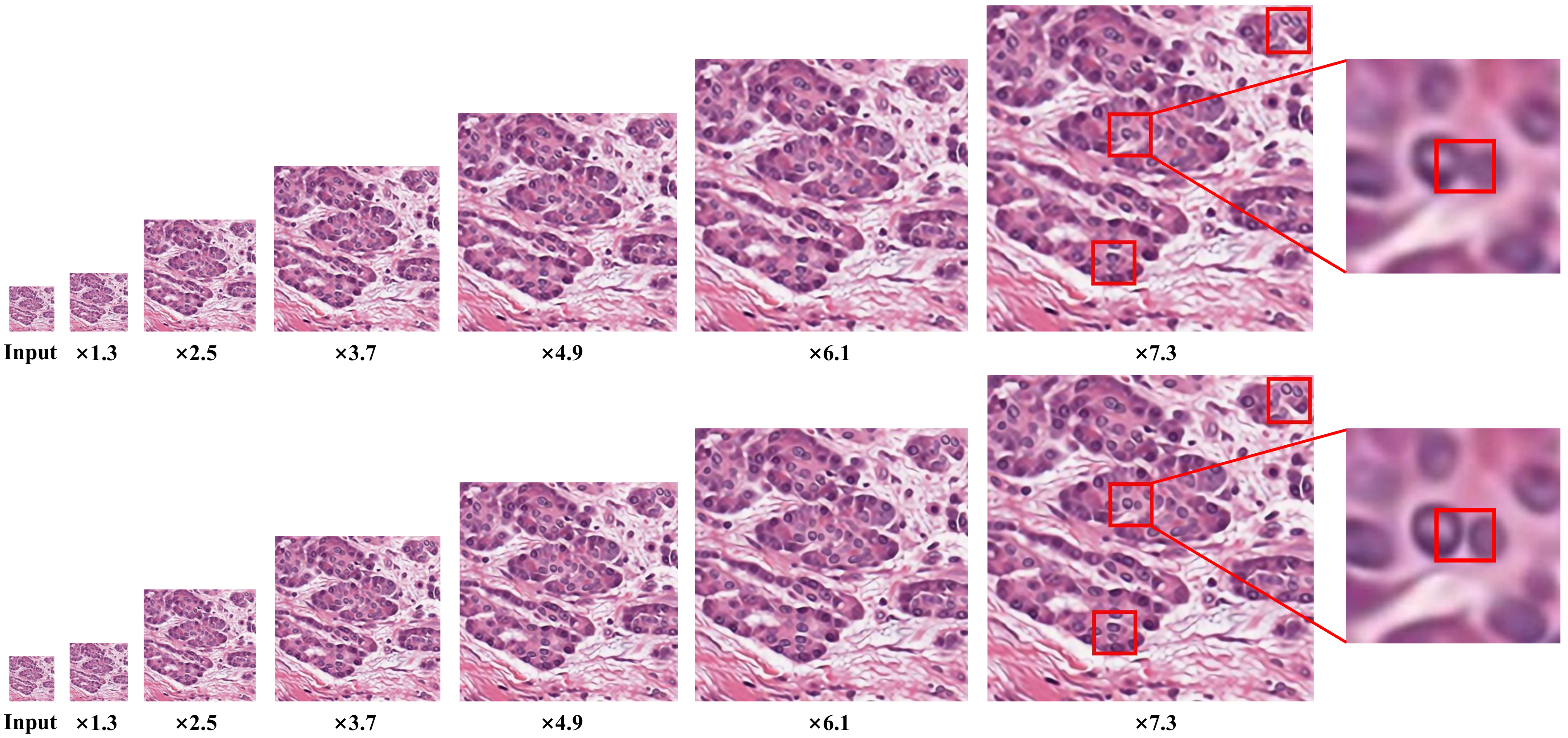}
\caption{Comparison of LIIF (upper row) and our ISTE (lower row) at non-integer scales.}
\label{fig6}
\end{figure}

\subsection{Ablation study}
To validate the effectiveness of each module in our proposed method, including the LFI, TL, STF, and LTD, we designed several variant networks for ablation experiments at scaling factors of ×2, ×3, and ×4 on the TCGA dataset.

\begin{table*}[h]
    \centering
    \caption{\centering{Ablation Study on the TCGA Dataset. The best results are indicated in \color[HTML]{000000} \textbf{bold}.}}
\setlength{\tabcolsep}{3pt}
\resizebox{1.0\columnwidth}{!}
    {
\begin{tabular}{|ccccccc|cc|cc|cc|}
\hline
\multicolumn{7}{|c|}{Model}                                                   & \multicolumn{2}{c|}{×2}                       & \multicolumn{2}{c|}{×3}                       & \multicolumn{2}{c|}{×4}                       \\ \hline
Dual-Branch & \multicolumn{1}{c|}{Single-Branch} & TL & LFI & STF & LTD & LPD & PSNR↑                & SSIM↑                  & PSNR↑                & SSIM↑                  & PSNR↑                & SSIM↑                  \\ \hline
×           & \multicolumn{1}{c|}{\checkmark}             & \checkmark  & ×   & ×   & \checkmark   & ×   & 37.45±1.041          & 0.9778±0.0053          & 32.02±0.910          & 0.9115±0.0163          & 29.14±0.866          & 0.8290±0.0248          \\ \hline
×           & \multicolumn{1}{c|}{\checkmark}             & ×  & \checkmark   & ×   & ×   & \checkmark   & 37.44±1.032          & 0.9778±0.0053          & 32.01±0.910          & 0.9115±0.0163          & 29.14±0.866          & 0.8290±0.0248          \\ \hline
\checkmark           & \multicolumn{1}{c|}{×}             & \checkmark  & \checkmark   & \checkmark   & ×   & \checkmark   & 37.63±1.041          & 0.9789±0.0052          & 32.04±0.912          & 0.9120±0.0163          & 29.17±0.867          & 0.8302±0.0248          \\ \hline
\checkmark           & \multicolumn{1}{c|}{×}             & \checkmark  & \checkmark   & ×   & \checkmark   & \checkmark   & 37.66±1.037          & 0.9791±0.0051          & 32.04±0.913          & 0.9121±0.0163          & 29.17±0.867          & 0.8301±0.0248          \\ \hline
\checkmark           & \multicolumn{1}{c|}{×}             & ×  & \checkmark   & \checkmark   & \checkmark   & \checkmark   & 37.64±1.039          & 0.9790±0.0051          & 32.04±0.913          & 0.9121±0.0163          & 29.17±0.867          & 0.8301±0.0248          \\ \hline
\checkmark           & \multicolumn{1}{c|}{×}             & \checkmark  & ×   & \checkmark   & \checkmark   & \checkmark   & 37.61±1.037          & 0.9788±0.0052          & 32.04±0.911          & 0.9121±0.0163          & 29.18±0.867          & 0.8303±0.0248          \\ \hline
\checkmark           & \multicolumn{1}{c|}{×}             & \checkmark  & \checkmark   & \checkmark   & \checkmark   & \checkmark   & \textbf{37.76±1.034} & \textbf{0.9796±0.0050} & \textbf{32.06±0.914} & \textbf{0.9124±0.0163} & \textbf{29.19±0.867} & \textbf{0.8307±0.0247} \\ \hline
\end{tabular}
    }
    \label{table4}
\end{table*}

\subsubsection{Evaluation of the local feature interactor}
For the features obtained from the encoder $F_{LR}$, the LFI enhances the interaction of features within local regions. To investigate the effectiveness of this module, we conducted an ablation experiment by directly removing the LFI from the ISTE framework. As shown in Table 4, all metrics are improved at all scaling factors using the LFI.

\subsubsection{Evaluation of the texture learner}
The TL is employed to enhance the learning of high-frequency textures in pathological images. To investigate the effectiveness of this module, we conducted an ablation experiment by replacing the module with a convolutional layer. As shown in Table 4, it can be seen that after ablating the TL, all metrics become worse at all scaling factors. To better illustrate the role of the TL, we visualized the features input to the TL and output from the TL, denoted as $F_{LR}$ and $F_{TL}$, respectively, in Fig.7. Compared to $F_{LR}$, the output feature map $F_{TL}$ from the TL contains richer texture information.

\begin{figure}[t]
\centering
\includegraphics[width=0.35\textwidth]{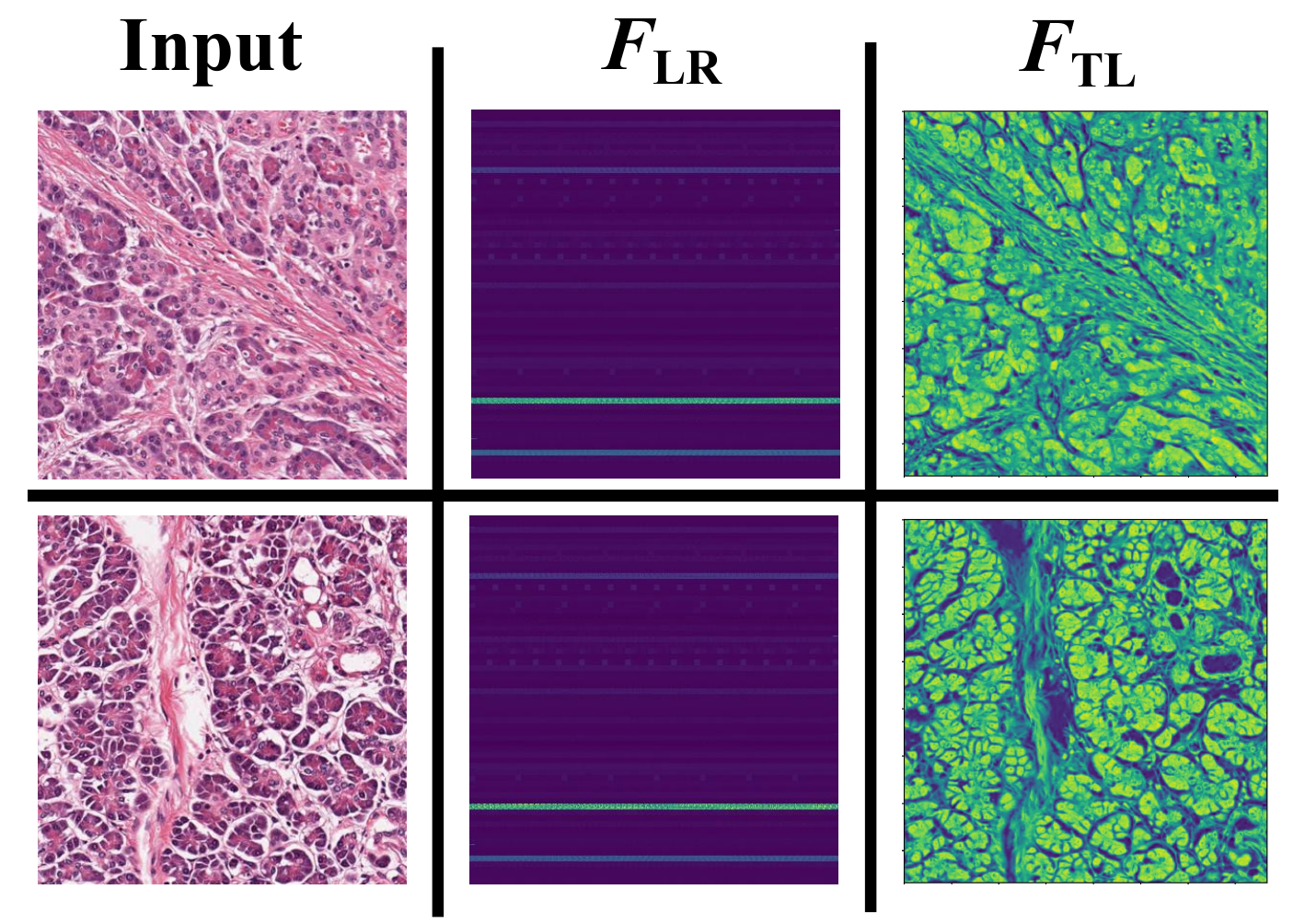}
\caption{Feature map visualization for the texture learner. $F_{LR}$ represents the feature map input to the texture learner and $F_{TL}$ represents the feature map output from the texture learner.}
\label{fig7}
\end{figure}

\subsubsection{Evaluation of the self-texture fusion module}
The STF module globally retrieves texture features that are most similar to $F_{LFIC}$ in $F_{TL}$ and fuses the retrieved features to $F_{LFIC}$. We designed a variant network without this module to evaluate its effectiveness. Specifically, we first take the feature $F_{LFIC}$ obtained from the feature aggregation branch of the framework and decode it directly through the LPD to obtain $I_{LPD}^{\prime}$. Then, we take the feature $F_{TL}$ obtained from the texture learning branch and decode it through the LTD to obtain $I_{LTD}^{\prime}$. We sum $I_{LPD}^{\prime}$ and $I_{LTD}^{\prime}$ to get the output of the variant network $I_{Pred}^{\prime}$. As shown in Table 4, all metrics become worse at all scaling factors after ablating the STF module. To illustrate the effectiveness of the STF module more intuitively, we visualized the path of the STF module to retrieve texture features on the TMA dataset in Fig.8. For the LR Patch during one training iteration, the starting point of the blue arrow is the position of the texture feature $F_{TL}$ retrieved by the STF module. The arrow points to the position where the feature $F_{LFIC}$ needs to be enhanced and fused with the retrieved texture feature $F_{TL}$. We visualize a proportion of the sampling pixels for a better demonstration in Fig.8. It can be seen that the STF module can effectively use similar tissue texture segments and cellular structure features in pathological images to assist reconstruction.

\begin{figure}[t!]
\centering
\includegraphics[width=0.4\textwidth]{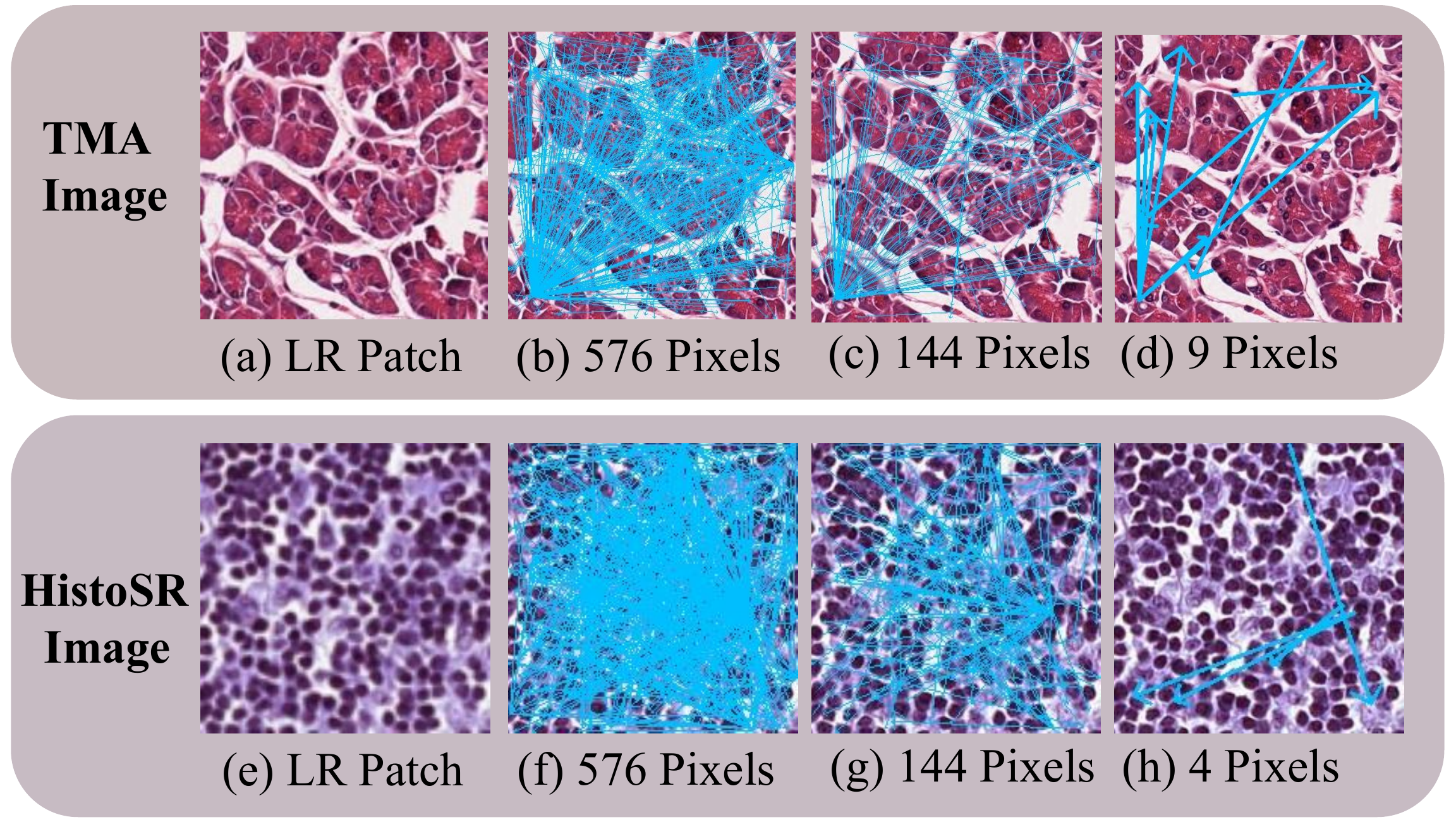}
\caption{Visualization of texture similarity retrieval for the STF module, where the blue arrow starting position indicates the position of the texture feature $F_{TL}$ retrieved by the STF module. The arrow points to the position where the pixel feature $F_{LFIC}$ needs to be enhanced and fused with the retrieved texture feature $F_{TL}$.}
\label{fig8}
\end{figure}

\subsubsection{Evaluation of the texture decoder for spatial domain-based enhancement}
The feature $F_{STF}$ is decoded into the pixel information $I_{LPD}$ in the spatial domain by the LPD. To accomplish spatial domain-based texture enhancement in the subsequent stage, LTD is employed to decode texture features acquired by the TL directly into spatial domain texture information $I_{LTD}$, and we sum $I_{LTD}$ with $I_{LPD}$ to obtain $I_{Pred}$. To demonstrate the effectiveness of the designed Spatial domain-based enhancement strategy, we removed the LTD in the ISTE framework and utilized only the pixels decoded by the LPD as the final prediction results. The results in Table 4 suggest that incorporating spatial domain-based texture enhancement can lead to improved results. To better illustrate the effectiveness of the spatial domain-based enhancement, we visualized the pixel information decoded by the LPD and the texture information decoded by the LTD in the framework of ISTE in Fig.9. It can be seen that the texture information $I_{LTD}$ decoded with LTD reveals clear outlines and texture features of the tissue cells and has more vibrant colors. This further illustrates the importance of LTD for spatial domain-based enhancement.

\begin{figure}[h]
\centering
\includegraphics[width=0.4\textwidth]{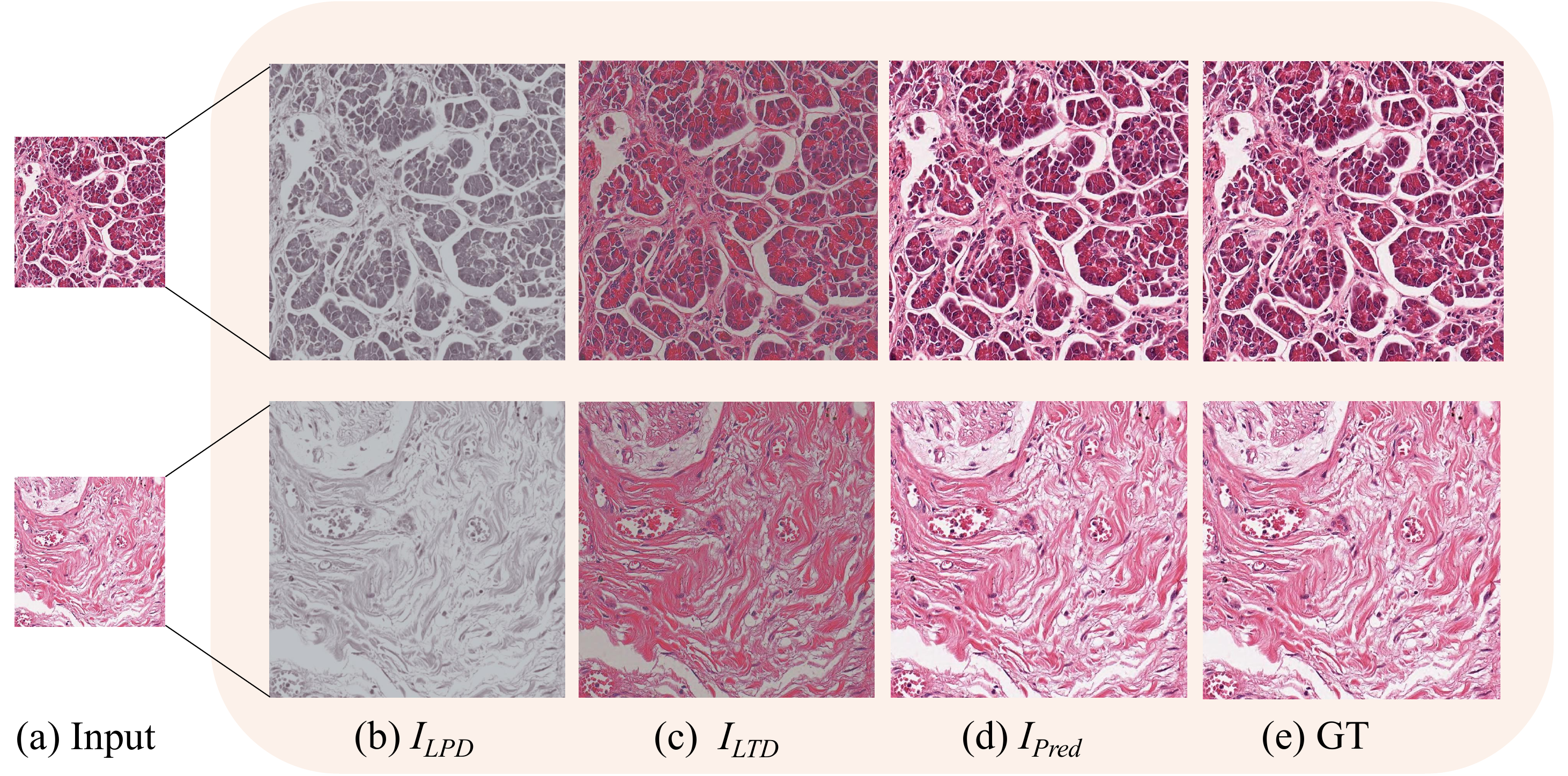}
\caption{(a) Input LR image; (b) Pixel information decoded by the LPD; (c) Texture information decoded by the LTD; (d) Output of the spatial domain-based enhancement; (e) Ground truth.}
\label{fig9}
\end{figure}

\subsubsection{Evaluation of the dual-branch architecture}
We designed two single-branch variant networks to evaluate the effectiveness of the proposed dual-branch architecture: (1) retaining only the TL and LTD in the ISTE framework and (2) retaining only the LFI and LPD in the ISTE framework. As shown in Table 4, the performance of the single-branch architecture is degraded compared to the dual-branch architecture.

\begin{table}[t]
\centering
\caption{Quantitative evaluation results of U-Net for gland segmentation on the GlaS dataset under different experimental settings.}
\label{table5}
\setlength{\tabcolsep}{3pt}
\scalebox{0.75}{
\begin{tabular}{c|cc|cc|cc}
\hline
\multirow{2}{*}{Experiment} & \multicolumn{2}{c|}{F1} & \multicolumn{2}{c|}{ObjDice} & \multicolumn{2}{c}{ObjHausdorff} \\ \cline{2-7} 
                            & Test A     & Test B     & Test A        & Test B       & Test A          & Test B         \\ \hline
Bicubic                     & 0.71       & 0.85       & 0.83          & 0.88         & 133.73          & 109.21         \\
HR U-Net                    & 0.84       & 0.88       & 0.89          & 0.92         & 100.57          & 84.64          \\
SISR                        & 0.92       & 0.93       & 0.94          & 0.95         & 77.74           & 65.81          \\
Original high resolution    & 0.95       & 0.93       & 0.96          & 0.96         & 66.70            & 61.17          \\ \hline
\end{tabular}}
\end{table}

\subsection{Downstream task experiments}
In this section, we experimentally demonstrate that the proposed SR method effectively enhances the performance of two downstream tasks: gland segmentation and malignancy classification. First, for the gland segmentation task, we trained and tested the state-of-the-art segmentation model U-Net \cite{ronneberger2015u} on the Glas dataset from the MICCAI 2015 Gland Segmentation Challenge \cite{sirinukunwattana2017gland}. The Glas dataset consists of a training set and two test sets, Test A and Test B. The training set contains 85 images and the corresponding labels, Test A contains 60 images and the corresponding labels, and Test B contains 20 images and the corresponding labels. We performed ×4 downsampling on HR images to generate LR images using bicubic interpolation. We compared segmentation results under the following settings: (1) Original high-resolution: Train U-Net on the original HR GlaS dataset for segmentation of original high-resolution images; (2) SISR: Directly employing U-Net trained on the original HR GlaS dataset for segmentation of the reconstructed images produced by our ISTE; (3) HR U-Net: Train U-Net on the reconstructed images produced by our ISTE for segmentation of original HR images; (4) Bicubic: Train U-Net on LR images obtained by bicubic interpolation for segmentation of original HR images. Table 5 shows the quantitative test results, where larger values indicate better performance for the F1 score and Object Dice score, while smaller values indicate better performance for object Hausdorff distance. It can be seen that the U-Net model trained on the reconstructed images of the SISR model performs better than the UNet model trained on the LR image dataset after bicubic interpolation, showing higher F1 scores and object Dice scores, as well as lower object Hausdorff distances. In particular, when tested on the Test B dataset, our results for segmentation of reconstructed images using U-Net trained on the original HR GlaS training set are close to those for segmentation of the original HR image, both with an F1 score of 0.93. Fig.10 shows representative results for different experimental setups, and it can be observed that U-Net trained on LR images produced the worst results; not only did it fail to detect small glands, but also the segmentation results of large glands appeared to be crippled. In contrast, the U-Net trained on the reconstructed image could outline the boundaries of the macro glands and detect the tiny glands. Compared to using LR images for training, using the generated SR images for training can improve the segmentation accuracy during testing.

\begin{figure}[h]
\centering
\includegraphics[width=0.4\textwidth]{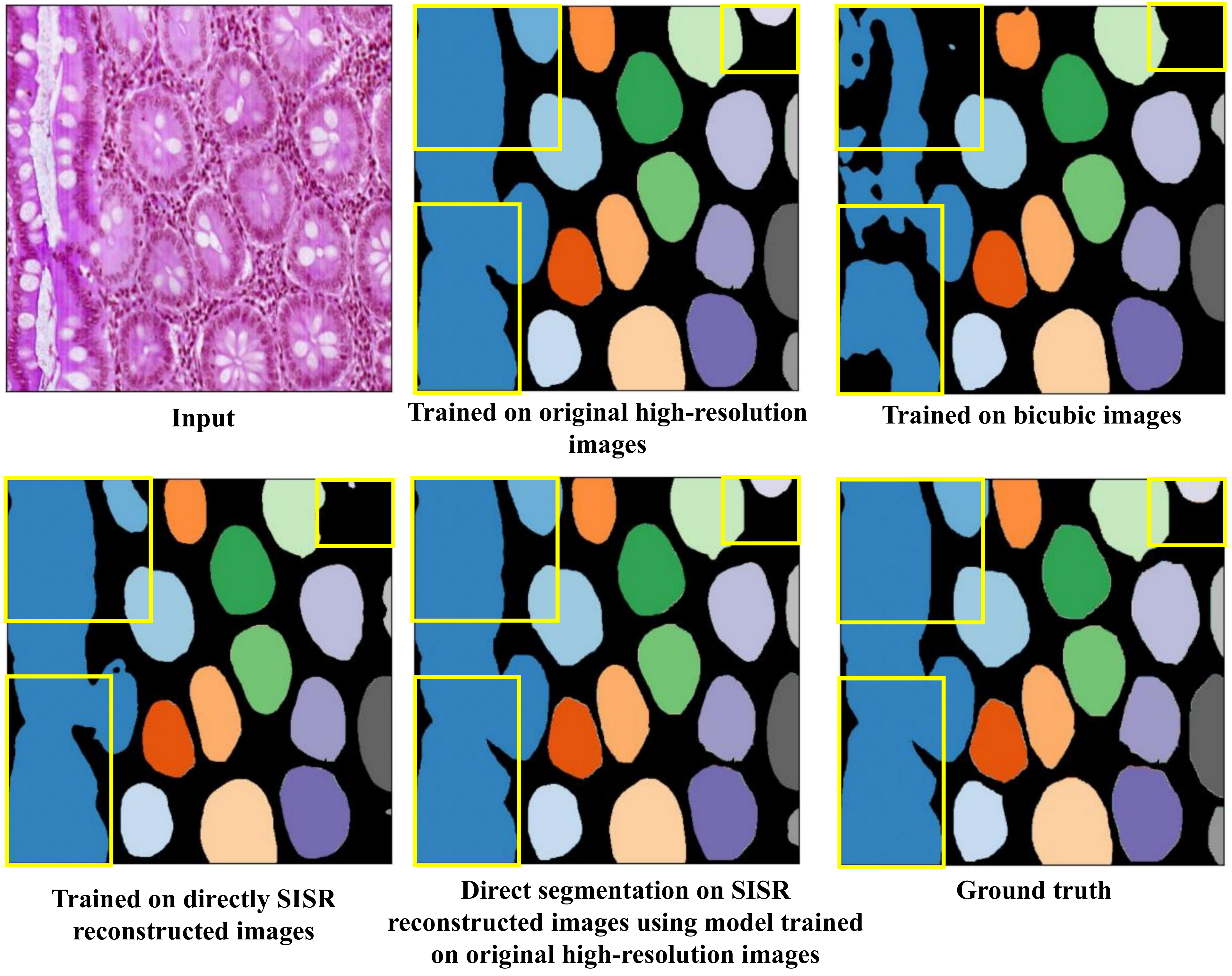}
\caption{Quantitative evaluation of UNet for gland segmentation on the GlaS dataset \cite{sirinukunwattana2017gland} with different experiment setups.}
\label{fig10}
\end{figure}

To further evaluate the contribution of the SR method to the malignancy classification task, we conducted tumor recognition on the PCam dataset \cite{veeling2018rotation}. The PCam dataset comprises 262,144 color images for training and 32,768 images for testing, with each image annotated with a binary label indicating the presence of metastatic tissue. We performed ×2 downsampling on HR images of the test set to generate LR images using bicubic interpolation. The ResNet-50 \cite{he2016deep} was chosen as the classifier and trained on the original PCam dataset. We compared classification results under the following settings: (1) Original: Directly employing trained ResNet-50 model to test on the original HR images in the test set; (2) Low resolution: Directly employing trained ResNet-50 model to test on the LR images of the test set; (3) Bicubic: Directly employing the trained ResNet-50 model to test on the bicubic interpolated images of the test set; (4) LIIF: Directly employing trained ResNet-50 model to test on the images generated by LIIF from the LR test set images; (5) ISTE: Directly employing trained ResNet-50 model to test on the images generated by our ISTE from the LR test set images. Table 6 illustrates the enhancement in diagnostic performance by the SR method. By introducing additional prior knowledge, our ISTE leads to a performance improvement, with an accuracy increase of 4.06\% compared to Bicubic. These results indicate that ISTE can improve classification performance by recovering more distinctive details.

\begin{table}[h]
\centering
\caption{The performance promotion using different SR methods in cancer detection.}
\label{table6}
\setlength{\tabcolsep}{3pt}
\scalebox{0.75}{
\begin{tabular}{c|c|c}
\hline
Experiment     & Accuracy                       & F1 score                      \\ \hline
Original       & 86.17\%                        & 0.8507                        \\ \hline
Low Resolution & 58.11\%                        & 0.2929                        \\
Bicubic        & 77.09\%                        & 0.7419                        \\
LIIF           & 80.54\%                        & 0.7721                        \\
ISTE(ours)           & {\color[HTML]{000000} \textbf{81.15\%}} & {\color[HTML]{000000} \textbf{0.7816}} \\ \hline
\end{tabular}}
\end{table}

\section{Conclusion}
In this work, we propose an innovative dual-branch framework ISTE based on implicit self-texture enhancement for arbitrary-scale histopathology image super-resolution. ISTE consists of a feature aggregation branch and a texture learning branch. We employ the feature aggregation branch to enhance the relevance of features in the local region while utilizing the texture learning branch to improve the learning of high-frequency texture details. We then design a two-stage texture enhancement strategy to fuse the features from the two branches to obtain SR images, where the first stage is feature-based texture enhancement and the second stage is spatial domain-based texture enhancement. Extensive experiments on publicly available datasets show that ISTE outperforms existing fixed-scale and arbitrary-scale SR methods across multiple scaling factors. Further experiments indicate that our method can enhance performance on two downstream tasks. In the future, we will continue to work on lightweight models and integrate the proposed SR models with existing diagnostic networks to improve diagnostic performance.

\section*{CRediT authorship contribution statement}
\textbf{Minghong Duan}: Writing – original draft, Software, Methodology, Investigation, Conceptualization. \textbf{Linhao Qu}: Writing – original draft, Validation, Supervision, Methodology, Data curation, Conceptualization. \textbf{Zhiwei Yang}: Validation, Software, Investigation. \textbf{Manning Wang}: Methodology, Supervision, Validation, Writing – review \& editing. \textbf{Chenxi Zhang}: Resources, Supervision, Validation, Writing – review \& editing. \textbf{Zhijian Song}: Resources, Validation, Writing – review \& editing.

\section*{Declaration of competing interest}
The authors declare that they have no known competing financial interests or personal relationships that could have appeared to influence the work reported in this paper.





\bibliographystyle{elsarticle-num}
\bibliography{ckwx}

\begin{thebibliography}{10}
\expandafter\ifx\csname url\endcsname\relax
  \def\url#1{\texttt{#1}}\fi
\expandafter\ifx\csname urlprefix\endcsname\relax\def\urlprefix{URL }\fi
\expandafter\ifx\csname href\endcsname\relax
  \def\href#1#2{#2} \def\path#1{#1}\fi

\bibitem{gilbertson2006primary}
J.~R. Gilbertson, J.~Ho, L.~Anthony, D.~M. Jukic, Y.~Yagi, A.~V. Parwani, Primary histologic diagnosis using automated whole slide imaging: a validation study, BMC clinical pathology 6 (2006) 1--19.

\bibitem{pantanowitz2011review}
L.~Pantanowitz, P.~N. Valenstein, A.~J. Evans, K.~J. Kaplan, J.~D. Pfeifer, D.~C. Wilbur, L.~C. Collins, T.~J. Colgan, Review of the current state of whole slide imaging in pathology, Journal of pathology informatics 2~(1) (2011) 36.

\bibitem{weinstein2004array}
R.~S. Weinstein, M.~R. Descour, C.~Liang, G.~Barker, K.~M. Scott, L.~Richter, E.~A. Krupinski, A.~K. Bhattacharyya, J.~R. Davis, A.~R. Graham, et~al., An array microscope for ultrarapid virtual slide processing and telepathology. design, fabrication, and validation study, Human pathology 35~(11) (2004) 1303--1314.

\bibitem{wilbur2011digital}
D.~C. Wilbur, Digital cytology: current state of the art and prospects for the future, Acta cytologica 55~(3) (2011) 227--238.

\bibitem{ghaznavi2013digital}
F.~Ghaznavi, A.~Evans, A.~Madabhushi, M.~Feldman, Digital imaging in pathology: whole-slide imaging and beyond, Annual Review of Pathology: Mechanisms of Disease 8 (2013) 331--359.

\bibitem{wu2023mmsrnet}
X.~Wu, Z.~Chen, C.~Peng, X.~Ye, Mmsrnet: Pathological image super-resolution by multi-task and multi-scale learning, Biomedical Signal Processing and Control 81 (2023) 104428.

\bibitem{nielsen2010virtual}
P.~S. Nielsen, J.~Lindebjerg, J.~Rasmussen, H.~Starklint, M.~Waldstr{\o}m, B.~Nielsen, Virtual microscopy: an evaluation of its validity and diagnostic performance in routine histologic diagnosis of skin tumors, Human pathology 41~(12) (2010) 1770--1776.

\bibitem{madabhushi2016image}
A.~Madabhushi, G.~Lee, Image analysis and machine learning in digital pathology: Challenges and opportunities, Medical image analysis 33 (2016) 170--175.

\bibitem{li2021single}
B.~Li, A.~Keikhosravi, A.~G. Loeffler, K.~W. Eliceiri, Single image super-resolution for whole slide image using convolutional neural networks and self-supervised color normalization, Medical Image Analysis 68 (2021) 101938.

\bibitem{lim2017enhanced}
B.~Lim, S.~Son, H.~Kim, S.~Nah, K.~Mu~Lee, Enhanced deep residual networks for single image super-resolution, in: Proceedings of the IEEE conference on computer vision and pattern recognition workshops, 2017, pp. 136--144.

\bibitem{mukherjee2018convolutional}
L.~Mukherjee, A.~Keikhosravi, D.~Bui, K.~W. Eliceiri, Convolutional neural networks for whole slide image superresolution, Biomedical optics express 9~(11) (2018) 5368--5386.

\bibitem{chen2020joint}
Z.~Chen, X.~Guo, C.~Yang, B.~Ibragimov, Y.~Yuan, Joint spatial-wavelet dual-stream network for super-resolution, in: Medical Image Computing and Computer Assisted Intervention--MICCAI 2020: 23rd International Conference, Lima, Peru, October 4--8, 2020, Proceedings, Part V 23, Springer, 2020, pp. 184--193.

\bibitem{sitzmann2020implicit}
V.~Sitzmann, J.~Martel, A.~Bergman, D.~Lindell, G.~Wetzstein, Implicit neural representations with periodic activation functions, Advances in neural information processing systems 33 (2020) 7462--7473.

\bibitem{tancik2020fourier}
M.~Tancik, P.~Srinivasan, B.~Mildenhall, S.~Fridovich-Keil, N.~Raghavan, U.~Singhal, R.~Ramamoorthi, J.~Barron, R.~Ng, Fourier features let networks learn high frequency functions in low dimensional domains, Advances in Neural Information Processing Systems 33 (2020) 7537--7547.

\bibitem{mildenhall2021nerf}
B.~Mildenhall, P.~P. Srinivasan, M.~Tancik, J.~T. Barron, R.~Ramamoorthi, R.~Ng, Nerf: Representing scenes as neural radiance fields for view synthesis, Communications of the ACM 65~(1) (2021) 99--106.

\bibitem{chen2021learning}
Y.~Chen, S.~Liu, X.~Wang, Learning continuous image representation with local implicit image function, in: Proceedings of the IEEE/CVF conference on computer vision and pattern recognition, 2021, pp. 8628--8638.

\bibitem{lee2022local}
J.~Lee, K.~H. Jin, Local texture estimator for implicit representation function, in: Proceedings of the IEEE/CVF conference on computer vision and pattern recognition, 2022, pp. 1929--1938.

\bibitem{upadhyay2019mixed}
U.~Upadhyay, S.~P. Awate, A mixed-supervision multilevel gan framework for image quality enhancement, in: International Conference on Medical Image Computing and Computer-Assisted Intervention, Springer, 2019, pp. 556--564.

\bibitem{juhong2023super}
A.~Juhong, B.~Li, C.-Y. Yao, C.-W. Yang, D.~W. Agnew, Y.~L. Lei, X.~Huang, W.~Piyawattanametha, Z.~Qiu, Super-resolution and segmentation deep learning for breast cancer histopathology image analysis, Biomedical Optics Express 14~(1) (2023) 18--36.

\bibitem{shahidi2021breast}
F.~Shahidi, Breast cancer histopathology image super-resolution using wide-attention gan with improved wasserstein gradient penalty and perceptual loss, IEEE Access 9 (2021) 32795--32809.

\bibitem{canny1986computational}
J.~Canny, A computational approach to edge detection, IEEE Transactions on pattern analysis and machine intelligence~(6) (1986) 679--698.

\bibitem{dong2014learning}
C.~Dong, C.~C. Loy, K.~He, X.~Tang, Learning a deep convolutional network for image super-resolution, in: Computer Vision--ECCV 2014: 13th European Conference, Zurich, Switzerland, September 6-12, 2014, Proceedings, Part IV 13, Springer, 2014, pp. 184--199.

\bibitem{zhang2018residual}
Y.~Zhang, Y.~Tian, Y.~Kong, B.~Zhong, Y.~Fu, Residual dense network for image super-resolution, in: Proceedings of the IEEE conference on computer vision and pattern recognition, 2018, pp. 2472--2481.

\bibitem{zhang2018image}
Y.~Zhang, K.~Li, K.~Li, L.~Wang, B.~Zhong, Y.~Fu, Image super-resolution using very deep residual channel attention networks, in: Proceedings of the European conference on computer vision (ECCV), 2018, pp. 286--301.

\bibitem{cavigelli2017cas}
L.~Cavigelli, P.~Hager, L.~Benini, Cas-cnn: A deep convolutional neural network for image compression artifact suppression, in: 2017 International Joint Conference on Neural Networks (IJCNN), IEEE, 2017, pp. 752--759.

\bibitem{kim2016accurate}
J.~Kim, J.~K. Lee, K.~M. Lee, Accurate image super-resolution using very deep convolutional networks, in: Proceedings of the IEEE conference on computer vision and pattern recognition, 2016, pp. 1646--1654.

\bibitem{wang2018esrgan}
X.~Wang, K.~Yu, S.~Wu, J.~Gu, Y.~Liu, C.~Dong, Y.~Qiao, C.~Change~Loy, Esrgan: Enhanced super-resolution generative adversarial networks, in: Proceedings of the European conference on computer vision (ECCV) workshops, 2018, pp. 0--0.

\bibitem{zhang2020residual}
Y.~Zhang, Y.~Tian, Y.~Kong, B.~Zhong, Y.~Fu, Residual dense network for image restoration, IEEE transactions on pattern analysis and machine intelligence 43~(7) (2020) 2480--2495.

\bibitem{chen2016trainable}
Y.~Chen, T.~Pock, Trainable nonlinear reaction diffusion: A flexible framework for fast and effective image restoration, IEEE transactions on pattern analysis and machine intelligence 39~(6) (2016) 1256--1272.

\bibitem{deng2021deep}
X.~Deng, Y.~Zhang, M.~Xu, S.~Gu, Y.~Duan, Deep coupled feedback network for joint exposure fusion and image super-resolution, IEEE Transactions on Image Processing 30 (2021) 3098--3112.

\bibitem{niu2020single}
B.~Niu, W.~Wen, W.~Ren, X.~Zhang, L.~Yang, S.~Wang, K.~Zhang, X.~Cao, H.~Shen, Single image super-resolution via a holistic attention network, in: Computer Vision--ECCV 2020: 16th European Conference, Glasgow, UK, August 23--28, 2020, Proceedings, Part XII 16, Springer, 2020, pp. 191--207.

\bibitem{chen2021pre}
H.~Chen, Y.~Wang, T.~Guo, C.~Xu, Y.~Deng, Z.~Liu, S.~Ma, C.~Xu, C.~Xu, W.~Gao, Pre-trained image processing transformer, in: Proceedings of the IEEE/CVF conference on computer vision and pattern recognition, 2021, pp. 12299--12310.

\bibitem{liang2021swinir}
J.~Liang, J.~Cao, G.~Sun, K.~Zhang, L.~Van~Gool, R.~Timofte, Swinir: Image restoration using swin transformer, in: Proceedings of the IEEE/CVF international conference on computer vision, 2021, pp. 1833--1844.

\bibitem{chen2023activating}
X.~Chen, X.~Wang, J.~Zhou, Y.~Qiao, C.~Dong, Activating more pixels in image super-resolution transformer, in: Proceedings of the IEEE/CVF Conference on Computer Vision and Pattern Recognition, 2023, pp. 22367--22377.

\bibitem{liu2018non}
D.~Liu, B.~Wen, Y.~Fan, C.~C. Loy, T.~S. Huang, Non-local recurrent network for image restoration, Advances in neural information processing systems 31 (2018).

\bibitem{mei2021image}
Y.~Mei, Y.~Fan, Y.~Zhou, Image super-resolution with non-local sparse attention, in: Proceedings of the IEEE/CVF Conference on Computer Vision and Pattern Recognition, 2021, pp. 3517--3526.

\bibitem{ma2021stsrnet}
J.~Ma, S.~Liu, S.~Cheng, R.~Chen, X.~Liu, L.~Chen, S.~Zeng, Stsrnet: Self-texture transfer super-resolution and refocusing network, IEEE Transactions on Medical Imaging 41~(2) (2021) 383--393.

\bibitem{zhang2019image}
Z.~Zhang, Z.~Wang, Z.~Lin, H.~Qi, Image super-resolution by neural texture transfer, in: Proceedings of the IEEE/CVF conference on computer vision and pattern recognition, 2019, pp. 7982--7991.

\bibitem{feng2021task}
C.-M. Feng, Y.~Yan, H.~Fu, L.~Chen, Y.~Xu, Task transformer network for joint mri reconstruction and super-resolution, in: Medical Image Computing and Computer Assisted Intervention--MICCAI 2021: 24th International Conference, Strasbourg, France, September 27--October 1, 2021, Proceedings, Part VI 24, Springer, 2021, pp. 307--317.

\bibitem{drifka2016highly}
C.~R. Drifka, A.~G. Loeffler, K.~Mathewson, A.~Keikhosravi, J.~C. Eickhoff, Y.~Liu, S.~M. Weber, W.~J. Kao, K.~W. Eliceiri, Highly aligned stromal collagen is a negative prognostic factor following pancreatic ductal adenocarcinoma resection, Oncotarget 7~(46) (2016) 76197.

\bibitem{drifka2015periductal}
C.~R. Drifka, J.~Tod, A.~G. Loeffler, Y.~Liu, G.~J. Thomas, K.~W. Eliceiri, W.~J. Kao, Periductal stromal collagen topology of pancreatic ductal adenocarcinoma differs from that of normal and chronic pancreatitis, Modern Pathology 28~(11) (2015) 1470--1480.

\bibitem{li2021dual}
B.~Li, Y.~Li, K.~W. Eliceiri, Dual-stream multiple instance learning network for whole slide image classification with self-supervised contrastive learning, in: Proceedings of the IEEE/CVF conference on computer vision and pattern recognition, 2021, pp. 14318--14328.

\bibitem{ronneberger2015u}
O.~Ronneberger, P.~Fischer, T.~Brox, U-net: Convolutional networks for biomedical image segmentation, in: Medical Image Computing and Computer-Assisted Intervention--MICCAI 2015: 18th International Conference, Munich, Germany, October 5-9, 2015, Proceedings, Part III 18, Springer, 2015, pp. 234--241.

\bibitem{sirinukunwattana2017gland}
K.~Sirinukunwattana, J.~P. Pluim, H.~Chen, X.~Qi, P.-A. Heng, Y.~B. Guo, L.~Y. Wang, B.~J. Matuszewski, E.~Bruni, U.~Sanchez, et~al., Gland segmentation in colon histology images: The glas challenge contest, Medical image analysis 35 (2017) 489--502.

\bibitem{veeling2018rotation}
B.~S. Veeling, J.~Linmans, J.~Winkens, T.~Cohen, M.~Welling, Rotation equivariant cnns for digital pathology, in: Medical Image Computing and Computer Assisted Intervention--MICCAI 2018: 21st International Conference, Granada, Spain, September 16-20, 2018, Proceedings, Part II 11, Springer, 2018, pp. 210--218.

\bibitem{he2016deep}
K.~He, X.~Zhang, S.~Ren, J.~Sun, Deep residual learning for image recognition, in: Proceedings of the IEEE conference on computer vision and pattern recognition, 2016, pp. 770--778.

\end{thebibliography}







\end{document}